\documentclass[fleqn,usenatbib]{mnras}

\usepackage[T1]{fontenc}

\DeclareRobustCommand{\VAN}[3]{#2}
\let\VANthebibliography\thebibliography
\def\thebibliography{\DeclareRobustCommand{\VAN}[3]{##3}\VANthebibliography}

\usepackage{graphicx}	
\usepackage{amsmath}	
\usepackage{amssymb}	

\usepackage{newtxtext,newtxmath}

\usepackage{tikz} 
\usetikzlibrary{positioning}
\usetikzlibrary{quotes,angles}
\usetikzlibrary{intersections}

\usepackage{pgfplots} 
\usepgfplotslibrary{fillbetween}

\usepackage{xcolor}

\title[Spectral Timing in Fairall 9]{Modelling Continuum Reverberation in AGN: A Spectral-Timing Analysis of the UV Variability Through X-ray Reverberation in Fairall 9}

\author[S. Hagen and C. Done]{
Scott Hagen,$^{1}$\thanks{E-mail: scott.hagen@durham.ac.uk}
Chris Done$^{1}$
\\
$^{1}$Department of Physics, University of Durham, South Road, Durham DH1 3LE, UK\\
}

\date{Accepted XXX. Received YYY; in original form ZZZ}

\pubyear{2022}

\pgfplotsset{compat=1.17} 
\begin{document}
\label{firstpage}
\pagerange{\pageref{firstpage}--\pageref{lastpage}}
\maketitle

\begin{abstract}
Continuum reverberation mapping of AGN can provide new insight into the nature and geometry of the accretion flow.
Some of the X-rays from the central corona irradiating the disc are absorbed, increasing the local disc temperature. This gives an additional re-processed contribution to the spectral energy distribution (SED) which is lagged and smeared relative to the driving X-ray light-curve. We directly calculate this reverberation from the accretion disc, creating fully time dependent SEDs for a given X-ray light-curve. We apply this to recent intensive monitoring data on Fairall 9, and find that it is not possible to 
produce the observed UV variability by X-ray reprocessing of the observed light-curve from the disc. Instead, we find that the majority of the variability must be intrinsic to the UV emission process, adding to evidence from changing-look AGN that this region has a structure which is quite unlike a Shakura-Sunyaev disc. We filter out this long timescale variability
and find that reprocessing alone is still insufficient to explain even the fast variability
in our assumed geometry of a central source illuminating a flat disc. The amplitude of reprocessing can be increased by 
any vertical structure such as the BLR and/or an inner disc wind, giving a better match. Fundamentally though the model is missing the major contributor to the variability, intrinsic to the UV/EUV emission rather than arising from reprocessing.

\end{abstract}




\defcitealias{Kubota18}{KD18}
\defcitealias{Gardner17}{GD17}

\section{Introduction}

Active galactic nuclei (AGN) are powered by accretion onto a supermassive black hole. This accretion flow is generally interpreted in the context of a \citet{Shakura73} disc, where energy dissipated through the flow gives rise to a radial temperature profile, increasing as the disc approaches the black hole. This leads to a spectral energy distribution (SED) composed of a sum of black-bodies, which for AGN peaks in the UV.

However, observations paint a more complex picture than a simple disc. Firstly, AGN spectra are always accompanied by a high energy X-ray tail \citep{Elvis94}
originating from the innermost regions of the accretion flow. Additionally there is also a soft X-ray excess, observed as an upturn below 1~keV which is remarkably constant in shape across a broad range of AGN (e.g \citealt{Porquet04, Gierlinski04}). Finally, the UV is often redder than expected from a standard disc, 
with a turnover again at a remarkably constant energy (e.g \citealt{Laor14}). Clearly then, AGN SEDs are more complex than a simple disc. 

One way to match the data is to assume that the accretion flow is radially stratifed, where the energy only thermalises to the standard \citet{Shakura73} blackbody temperature at large radii, and inside of this the accretion energy instead emerges as warm or hot Comptonised emission  \citep{Done12, Kubota18}. This is designed to allow the models enough flexibility to match the broadband SED, with enough constraints from energetics to fit the data without too much degeneracy. 

We can use spectral-timing to test this idea of a radial transition in the flow. Observations of AGN display strong X-ray variability, which can be used to constrain the nature and geometry of the accretion flow (e.g \citealt{Uttley14}). In particular, illumination of the disc (blackbody or warm Compton emitting) by the fast variable hot Compton X-ray flux leads to a fast variable reprocessed component, which correlates with but lags behind the X-rays by a light travel time. This reverberation mapping was originally proposed by \citet{Blandford82} as a means to measure the size scale of the broad emission line region (BLR) in AGN. The lines respond to changes in the ionising (UV to soft X-ray) lightcurve in a way that can be described as a convolution of the driving lightcurve with a transfer function which contains all the light travel time delays from the specific geometry. Observing the driving lightcurve and its delayed and smoothed emission line response gives information on the size scale and geometry of the BLR. This is often simply condensed to a single number, which is the mean lag  revealed in cross-correlation (e.g \citealt{Welsh91, Peterson93, Horne04, Peterson04b}).

However, the technique of reverberation mapping is quite general, and can be used to map continuum components as well as emission lines. 
Recently, there has been a drive for intensive multi-wavelength monitoring campaigns of AGN to map the accretion disc geometry from observations of the reprocessed continuum emission produced by the fast variable X-ray source illuminating the disc. 
(e.g \citealt{Edelson15, Edelson19, McHardy14, McHardy18}). In particular, the use of space telescopes (especially \textit{Swift}) in these campaigns has allowed high-quality and near continuous monitoring over extended periods of time, with simultaneous data of both the (assumed) driving hard X-ray and the disc reprocessed UV/optical emission. 

The results from these continuum reverberation campaigns are very surprising. In general, the UV variability is not well correlated with the X-ray variability which is meant to be its main driver. Instead, all wavebands longer than the far UV correlate well with the far UV lightcurve, but on a timescale which is longer than expected from a standard disc (e.g. the compilation of \citealt{Edelson19}). 

Here, instead of working backwards to a geometry from cross-correlation time lags, we work forward from a geometry given by the new SED models. Crucially, as well as predicting the smoothing/lag from light travel time delays, this also allows us to predict the {\em amplitude} of the response, giving a predicted UV lightcurve which can be compared point by point to the observed UV lightcurve. 

This approach was first used by 
 \citet{Gardner17} (hereafter \citetalias{Gardner17}) to model the light-curves in NGC 5548, but here the strong extrinsic X-ray variability from an unusual obscuration event made comparison difficult (e.g. \citealt{Mehdipour16, Dehghanian19a}). NGC 4151 gave a cleaner comparison as although this also shows strongly variable absorption, it is bright enough to be monitored by the Swift BAT instrument, sensitive to the higher energy 20-50~keV flux which is unaffected by the absorption. This showed clearly that there was a radial transition in the flow, with no reverberating material within $500-1000R_g$ \citep{Mahmoud20}. Such a hole in the optically thick material in this object is also now consistent with the X-ray iron line profile \citep{Miller18} and its  reverberation \citep{Zoghbi19}, despite previous claims to the contrary (e.g. \citealt{Keck15, Zoghbi12, Cackett14}).
 The new SED models had indeed predicted truncation of the optically thick disc material, though on somewhat smaller size scales of $50-100R_g$. The size of the response also showed that there was substantial contribution to the reprocessed flux from above the disc plane. This is almost certainly the same material as is seen in the variable absorption, which is clearly a wind launched on the inner edge of the BLR (\citealt{Kaastra14, Dehghanian19b, Chelouche19}). An additional diffuse UV contribution from X-ray illumination of this wind/BLR \citep{Korista01, Lawther18} means that the 
 SED model fits to the data overestimated the UV from the accretion flow itself, which probably led to the underestimate of the truncation radius in NGC4151. 

In this paper we follow the approach of \citetalias{Gardner17} and \citet{Mahmoud20}, but for Fairall 9 (hereafter F9). This explores a very different part of parameter space in terms of mass accretion rate.
Both NGC 4151 and NGC 5548 are at $0.01-0.03L_{\text{Edd}}$, close to the changing state transition so the disc should be quite strongly truncated, with a hot flow replacing the inner disc \citep{Noda18,Ruan19}. 
By contrast, F9 has $L\sim 0.1L_{\text{Edd}}$, so should have much more inner disc. F9 is also most likely an almost face on AGN as it shows very little 
line of sight absorption from either cold 
or ionised material (bare AGN: \citealt{Patrick11}). This means that the X-ray lightcurve is more likely representative of the intrinsic variability, rather than being heavily contaminated by extrinsic absorption variability (though we note there are occasional obscuration dips: \citealt{Lohfink16}). 

We construct a full spectral-timing code, {\sc agnvar},
which predicts variability at any wavelength from the new SED models ({\sc agnsed} in {\sc xspec}). We make this publicly available as a {\sc python} module\footnote{\url{https://github.com/scotthgn/AGNvar}}, and apply it to the recent intensive monitoring data on F9 \citep{Hernandez20}.

We describe the model in section 2. The underlying SED is necessary to understand the overall energetics: a source where the X-ray luminosity is as large as the UV luminosity can give a much stronger UV response to a factor 2 change in X-ray flux than a source like F9 where the X-ray power is 10x smaller than the UV (\citealt{Kubota18}). The radial size scale of the transition regions in the disc is likewise set by the SED fits, which determines the response light travel time. We outline our method for evolving the SED along with the light-curve, and explore how our model system responds to changes in X-ray illumination. We apply this to the mean SED of F9, and form the full time-dependent lightcurves from reprocessing of the observed X-rays in  Section 4. Our model fails to describe the data, most clearly as it produces much less variability amplitude than observed for any reasonable scale height of the X-ray source. This is clearly a consequence of simple energy arguments from the SED as the UV luminosity is considerably larger than the X-ray luminosity, so even a factor 2 change in X-ray flux has only limited impact on the UV, especial in the geometry assumed here of a central source illuminating a flat, truncated disc.
This clearly shows that most of the variability in the UV is intrinsic (assuming the observed X-rays are isotropic), which is at odds with standard Shakura-Sunyaev disc models, as these should only be able to vary on a viscous timescale (e.g. \citealt{Lawrence18}). Our work highlights the lack of understanding of the structures which emit most of the accretion power in AGN.

\section{Modelling the Response of the Accretion Flow}

Throughout our analysis we fix the black hole mass and distance to $M=2 \cdot 10^{8}\,M_{\odot}$, $d = 200$\,Mpc \citep{Bentz15}, and assume an inclination angle of $\cos(i) = 0.9$. We will also adopt the standard notation for radii, where $r$ is the dimensionless gravitational radius, and $R$ is the physical distance from the black hole. These are related by $R = rR_{G}$, where $R_{G} = GM/c^{2}$. 

\subsection{The Intrinsic SED}

We start by considering the intrinsic disc emission. We will follow the {\sc agnsed} model described in \citet{Kubota18} (hereafter \citetalias{Kubota18}) , and give a brief summary here.

The accretion flow is assumed to be radially stratified, but with a standard Novikov-Thorne emissivity profile $\epsilon_{NT}(R)$ \citep{Novikov73, Page74}. 
We divide the flow into annuli of width $\Delta R$, 
which emits luminosity $L(R) \Delta R 
= 2\times 2\pi R \epsilon_{NT} \Delta R$
(where the factor 2 is for each side of the disc).
Each disc annulus emits as a blackbody for $R_{\text{out}}<R<R_{w}$, with temperature $T_{NT} = (\epsilon_{NT}/\sigma_{SB})^{1/4}$ 
(\citealt{Shakura73}). 

Inwards of this, for $R_{w}<R<R_{h}$, thermalisation is incomplete, leading to the luminosity instead being emitted as warm, optically thick Comptonisation.
Following \citetalias{Kubota18}, we assume that this warm comptonisation region overlies a passive (non-emitting) disc (\citealt{Petrucci18}), so its seed photons are set by reprocessing the warm Comptonisation power on the passive disc material. Unlike the models of \cite{Petrucci18}, this sets the seed photon temperature self consistently from the size scale, which is just the same as the expected 
disc temperature, $T_{NT}(R)$. This means that the seed photon temperature changes in the warm Comptonisation region from $R_{w}$ to $R_{h}$, again unlike the \citet{Petrucci18} models, where it is only a single temperature blackbody. 

We model the warm Compton spectrum using 
{\sc nthcomp} \citep{Zdziarski96, Zycki99}, normalising the output luminosity of each annulus to the intrinsic disc luminosity of the annulus. The seed photon temperature is set as above, leaving two additional parameters: the electron temperature, $kT_{e, w}$, setting the high-energy roll over of the resulting spectral component, and the photon index $\Gamma_{w}$, which determines the spectral slope. Reprocessing of optically thick, warm Comptonisation from a passive disc gives $\Gamma_{w}=2.5-2.7$
\citep{Petrucci18}.

This still does not explain the full SED, as AGN spectra are always accompanied by a high energy X-ray tail (e.g \citealt{Elvis94}). As in \citetalias{Kubota18}, we consider this emission originates from the innermost accretion flow, where the disc has evaporated into an optically thin, geometrically thick corona (e.g \citealt{Narayan95, Zdziarski99, Liu99, Rozanska00b}) extending down to the innermost stable circular orbit, $R_{\text{isco}}$. We still assume that this is heated by the 
Novikov-Thorne emissivity, giving $L_{\text{diss}}$, but with seed photons from the inner edge of the warm Comptonisation region, so that the total X-ray luminosity is increased by this seed photon contribution giving total power $L_{x} = L_{\text{diss}} + L_{\text{seed}}$ (see \citetalias{Kubota18} for details on calculating these). The model assumes that the inner disc is replaced by the hot Comptonising plasma, so unlike the warm Comptonisation region, there is no underlying disc to provide a source for these seed photons. Instead, we assume that the seed photons originate from the truncated disc beyond $R_h$. We assume the seed photons come mainly from the inner edge of the truncated disc, 
so if there is a warm Comptonisation region the seed temperature is $T_{NT}(R_{h}) \exp(y_{w})$, where $y_{w}$ is the Compton y-parameter for the warm Comptonisation region. If there is no warm Comptonisation region, then the seed temperature is simply the temperature of the inner disc, $T_{NT}(R_{h})$. Again, like the warm region, we model this with {\sc nthcomp} and leave the electron temperature, $kT_{e, h}$, and photon index, $\Gamma_{h}$, as parameters within the code. This sets the spectral shape of the warm Compton region, which we then normalise to the total luminosity $L_{x}$.

We now have a model, identical to the one given in \citetalias{Kubota18}, consisting of three regions, with absolute size scale set by the black hole mass via $R_g$. A standard outer disc, located between $r_{\text{out}}$ and $r_{w}$, a warm Comptonisation region, where the disc emission fails to thermalize, located between $r_{w}$ and $r_{h}$, and a hot Comptonizing corona replacing the optically thick disc from $r_{h}$ to $r_{\text{isco}}$. This geometry is sketched in Fig. \ref{fig:mod_pics}a. The total SED is then the total contribution from each region added together, and forms the basis of our spectral timing model. 	

\subsection{Contribution from Reprocessing}

\begin{figure}
    \centering
    \begin{tikzpicture}
    
    
    \filldraw[color=blue] (0, 0) circle (25pt);
    \filldraw[color=white] (0, 0) circle (8pt) ;
    \filldraw[color=black] (0, 0) circle (4pt) ;
    
    \filldraw[color=green] (-0.89, 0.08) -- (-3, 0.15) -- (-3, -0.15) -- (-0.89, -0.08) ;
    \filldraw[color=red] (-3, 0.15) -- (-5, 0.2) -- (-5, -0.2) -- (-3, -0.15) ;
    
    \node (rout) at (-5.1, 1.45) {\large{$r_{\text{out}}$}} ;
    \node (rw) at (-3.1, 1.45) {\large{$r_{w}$}} ;
    \node (rh) at (-1, 1.45) {\large{$r_{h}$}} ;
    \node (risc) at (-0.1, 1.45) {\large{$r_{\text{isco}}$}} ;
    
    \draw (-5, 0) -- (-5, 1.3) ;
    \draw (-3, 0) -- (-3, 1.3) ;
    \draw (-0.89, 0) -- (-0.89, 1.3) ;
    \draw (-0.3, 0) -- (-0.3, 1.3) ;
    
    \node (I) at (2, 0) {\large{\textbf{(a.)}}} ;

    
    \filldraw[color=black] (0, -4) circle (4pt) ;
    \filldraw[color=blue] (0, -2.5) circle (2pt) ;
    
    \filldraw[color=green] (-0.89, -3.92) -- (-3, -3.85) -- (-3, -4.15) -- (-0.89, -4.08) ;
    \filldraw[color=red] (-3, -3.85) -- (-5, -3.8) -- (-5, -4.2) -- (-3, -4.15) ;
    
    \draw[<->] (0.2, -4) -- (0.2, -2.5) ;
    \node (hx) at (0.45, -3.3) {\large{$h_{x}$}} ;
    \node (Lx) at (0.2, -2.3) {\large{$L_{x}$}} ;
    
    \draw[<->] (0, -4.4) -- (-3.5, -4.4) ;
    \node (rad) at (-1.6, -4.6) {\large{$r$}} ;
    
    \draw[->] (0, -2.5) coordinate(LX) -- (-3.5, -3.8) coordinate(D) ;
    \draw[->] (-3.5, -3.8) -- (-3.5, -3) coordinate(N) ;
    
    \pic [draw, <->, "$n$", angle eccentricity=1.5] {angle=LX--D--N} ;
    
    \node (II) at (2, -4) {\large{\textbf{(b.)}}} ;
    
    \end{tikzpicture}
    \caption{\textit{(a)}: Schematic of the model geometry considered in this paper. Between $r_{\text{out}}$ and $r_{w}$ we have a standard accretion disc (red region), emitting like a multi-colour black-body. From $r_{w}$ to $r_{h}$ we have the warm Comptonisation region (green), where the disc fails to thermalize leading to Comptonisation of the underlying disc photons. Finally, between $r_{h}$ and $r_{\text{isco}}$ the disc has evaporated into the hot Comptonisation region, which we consider as a spherical corona (blue). \textit{(b)}: Schematic of the geometry we use to calculate the re-processed emission. Here we approximate the spherical corona as a point source located a height $h_{x}$ above the spin axis; following \citetalias{Gardner17}.}
    \label{fig:mod_pics}
\end{figure}
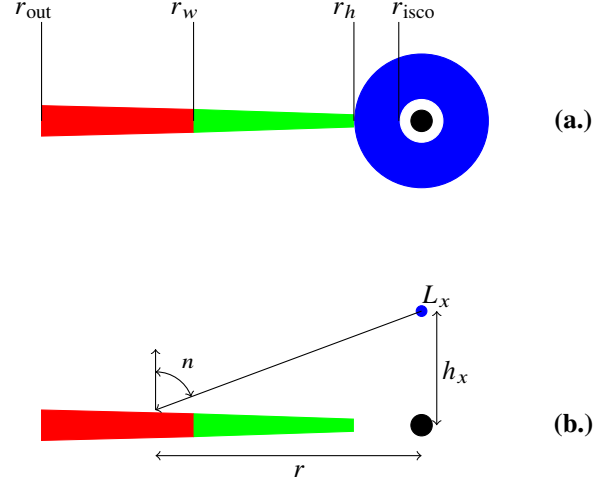

So far we have only considered the intrinsic emission from the accretion flow. However, a portion of the photons emitted by the corona will be incident on the disc, and a fraction of these will be absorbed and re-emitted. This will give a contribution to the local temperature at a point on the disc $\propto F_{\text{rep}}^{1/4}$, which is dependent on both the geometry of the disc (e.g \citealt{Zycki99, Hartnoll00}), and the corona. In our current picture we consider the inner corona to be an extended sphere, with luminosity per unit volume which depends on radius. The flux at a given point on the disc from illumination requires integrating the diffuse emission over the entire corona. Repeating this for the entire disc is clearly computationally expensive. However, \citetalias{Gardner17} showed that there is little  difference in the illumination pattern between a radially extended corona powered by Novikov-Thorne emissivity and a point source located a height $h_{x}\sim 10$ above the black hole. This removes the need for the expensive integration, and also makes the calculation of time delays in the next section much simpler, giving the picture illustrated in Fig. \ref{fig:mod_pics}b. The flux seen by a point on a disc then takes the simple form \citep{Zycki99}

\begin{equation}
    F_{\text{rep}}(r) = \frac{L_{x} \cos(n)}{4 \pi (r^{2} + h_{x}^{2})R_{G}^{2}}
    \label{eqn:Frep}
\end{equation}
where $\cos(n) = h_{x}/(r^{2} + h_{x}^{2})^{1/2}$ is the angle between the incident ray on the disc and the surface normal, and $L_{x}$ is the coronal luminosity. The effective temperature at a given radius will then be
\begin{equation}
    \sigma T_{\text{eff}}^{4}(r) = \sigma T_{NT}^{4}(r) + F_{\text{rep}}(r) (1 - A)
    \label{eqn:Teff}
\end{equation}
where $\sigma$ is the Stefan-Boltzman constant, and $A$ is the disc albedo. As in \citetalias{Kubota18} we fix the albedo to 0.3. We note that unlike \citetalias{Kubota18} we use $L_{x}$ rather than $0.5\,L_{x}$ for the luminosity seen by the disc. This is because we allow the disc to be heated from both sides. Due to symmetry, this is the same as using only one side for the geometry but letting it see the full X-ray luminosity.

\subsection{Time-dependent Reprocessing}
We now extend the re-processed contribution to the SED into a time-dependent SED model by considering the light-travel time between the X-ray source, accretion disc, and observer as in 
 \citetalias{Gardner17} and \citet{Mahmoud20}.

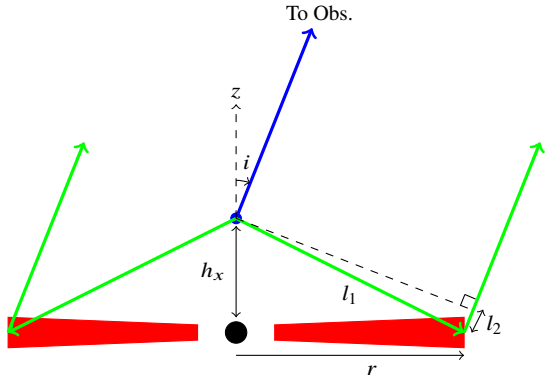
\begin{figure}
    \centering
    \begin{tikzpicture}
    
        \filldraw[color=black] (0, 0) circle (4pt) ;
        \filldraw[color=red] (-0.5, 0.1) -- (-3, 0.2) -- (-3, -0.2) -- (-0.5, -0.1) ;
        \filldraw[color=red] (0.5, 0.1) -- (3, 0.2) -- (3, -0.2) -- (0.5, -0.1) ;
        
        \filldraw[color=blue] (0, 1.5) circle (2pt) ;
        
        \draw[<->, thin] (0, 0.2) -- (0, 1.4) ;
        \draw[->, very thick, color=blue] (0, 1.5) -- (1, 4) coordinate (Obs) ;
        
        \draw[->, very thick, color=green] (0, 1.5) -- (3, 0) ;
        \draw[->, very thick, color=green] (3, 0) -- (4, 2.5) ;
        
        \draw[->, very thick, color=green] (0, 1.5) -- (-3, 0) ;
        \draw[->, very thick, color=green] (-3, 0) -- (-2, 2.5) ;
        
        \draw[dashed] (0, 1.5) -- (3.1, 0.3) ;
        \draw (2.95, 0.35) -- (3.01, 0.5) ; 
        \draw (3.01, 0.5) -- (3.15, 0.44) ;
        
        \node (l1) at (1.5, 0.5) {$l_{1}$} ;
        \draw[<->] (3.1, 0) -- (3.25, 0.3) ;
        \node (l2) at (3.4, 0.1) {$l_{2}$} ;
        
        \node (hx) at (-0.3, 0.8) {$h_{x}$} ;
        \draw[->] (0, -0.3) -- (3, -0.3) ;
        \node (r) at (1.8, -0.5) {$r$} ;
        
        \draw [dashed, ->] (0, 1.5) coordinate (Lx) -- (0, 3) coordinate (Z) ;
        \node (z) at (0, 3.15) {$z$} ;
        \pic[draw, <-, "$i$", angle eccentricity = 1.5] {angle=Obs--Lx--Z} ;
        
        \node (obs) at (1.1, 4.2) {To Obs.} ;
        
    \end{tikzpicture}
    \caption{Illustration of the geometry considered when calculating the time delay between the direct and re-processed emission. For simplicity we use a lamppost geometry. The direct coronal emission is shown by the blue line, while the indirect emission via the disc is given in green. The labels $l_{1}$ and $l_{2}$ indicate the portion of the indirect travel path that contribute to the time-delay.}
    \label{fig:delay_geom}
\end{figure}

 Here we follow the method in \citet{Welsh91}. The direct coronal emission will have a shorter path length to the observer, than the re-processed emission that first has to travel via the disc. This is illustrated in Fig. \ref{fig:delay_geom}, where we can see that the indirect emission has a travel path increased by $l_{1} + l_{2}$ with respect to the direct emission, which leads to the indirect emission lagging behind the direct by a time delay $\tau = (l_{1} + l_{2})/c$. Clearly $l_{1}$ and $l_{2}$ depend on the disc position being considered, hence we can re-write our path difference in terms of the disc coordinates $r$ and $\phi$. The result is given in Eqn. \ref{eqn:tau}, which is similar to that given in \citet{Welsh91}, however with the addition of the coronal height $h_{x}$. 

\begin{equation}
    \tau(r, \phi) = \frac{R_{G}}{c} \bigg\{ \sqrt{r^{2} + h_{x}^{2}} + h_{x}\cos(i) - r\cos(\phi)\sin(i) \bigg\}
    \label{eqn:tau}
\end{equation}

We define a grid across the accretion disc, with default spacing $d\log r = 0.01$ and $d\phi = 0.01$\,rad, and use Eqn. \ref{eqn:tau} to construct time-delay srufaces; or isodelay surfaces. These delay surfaces are used to map an observed X-ray light-curve $F_{x, \text{obs}}(t)$ onto the disc. For each time $t$ within the light-curve, a point $(r, \phi)$ on the disc will see the X-ray luminosity from time $t - \tau$. Assuming the X-ray coronal luminosity varies exactly like the observed light-curve, such that $L_{x}(t)/\langle L_{x}(t) \rangle = F_{x, \text{obs}}/\langle F_{x, \text{obs}} \rangle$, then we have that the disc temperature will vary as
\begin{equation}
    \sigma T_{\text{eff}}^{4}(r, \phi, t) = \sigma T_{NT}^{4}(r) + \frac{\cos(n) (1-A)}{4 \pi (r^{2} + h_{x}^{2}) R_{G}^{2}} L_{x}(t - \tau(r, \phi))
    \label{eqn:Teff_t}
\end{equation}

To calculate the time-dependent SEDs, we start by calculating the disc temperature within each grid-point across the disc, for every time-step in the input light-curve. In the interest of computational efficiency, and since $\Delta R << c\Delta t$ across the extent of our accretion disc, we flatten the 2D grid into a radial grid by calculating the mean temperature within each annulus; based off the grid points within that annulus. The SED for each time-step is then calculated following section 2, resulting in a series of model SEDs covering the duration of the light-curve.

We then extract lightcurves in a given energy band by 
defining a midpoint energy, $E_{\text{mid}}$, and a bin-width, $dE$. The model light-curve is simply the integral of the SED flux-density within the energy bin centred on $E_{\text{mid}}$ at each time-step $t$. 
This can be directly compared to the observed fluxes.
However, we choose instead to show light-curves in terms of the mean normalised flux $F/\langle F \rangle$ so that both data and model are dimensionless.

This all assumes that the effects of general relativity are small, unlike the code of \citet{Dovciak22} as used for reverberation by \citet{Kammoun19,Kammoun21a}. A fully general relativistic treatment is required for very small corona height, however \citet{Kammoun21a} show that these corrections are negligible for a large coronal height.

\subsection{Understanding the Disc Response}

\begin{figure*}
    \centering
    \includegraphics[width=\textwidth]{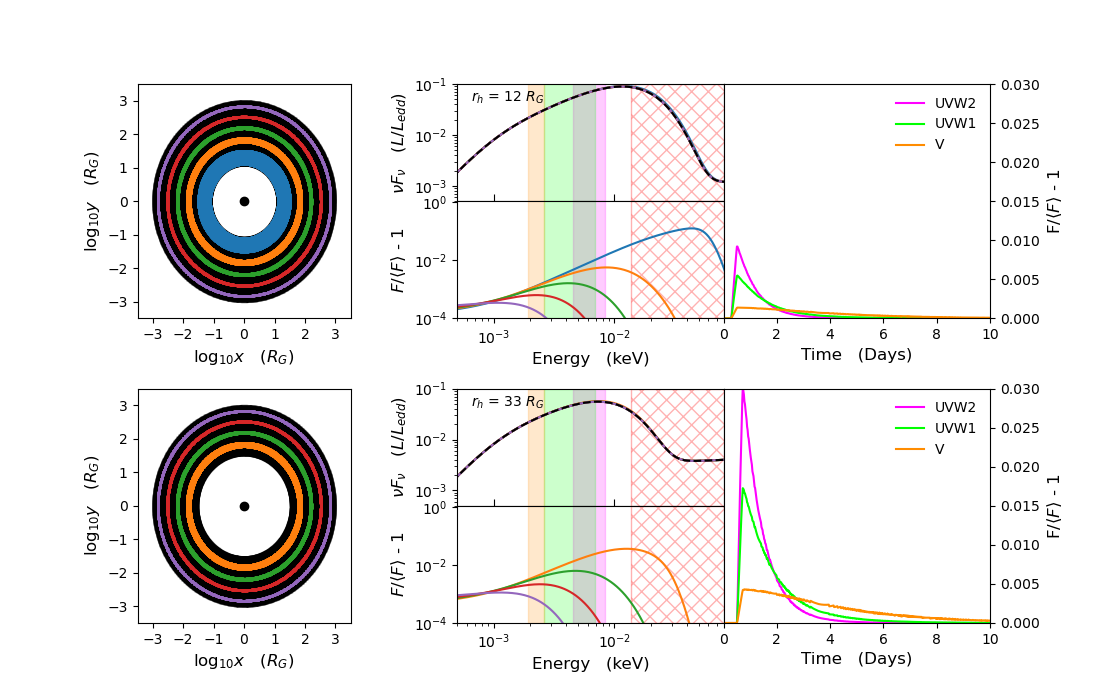}
    \caption{SED snapshots as an X-ray flash propagates across an accretion disc, with truncation radius $r_{h} = 12$ and $33$ (top and bottom rows respectively). The left column shows the position of the X-ray flash, as seen by the disc, at each time (after the initial flash) we extract SED snapshots for. These times are 0.5 days (blue), 1 day (orange), 2 days (green), 4 days (red), and 8 days (purple). The middle column shows (top panel) each SED snapshot overlaid on top of the mean SED (black dashed line) and (bottom panel) the residual of the SED snapshot with respect to the mean. The red crosshatch region in this column indicates the unobservable part of the spectrum. The rightmost column shows the responses, extracted for UVW2 (magenta), UVW1 (lime green), and V (orange) bands (assuming the same width as the \textit{Swift}-UVOT filters). These bandpasses are also highlighted in the middle column as the shaded coloured regions, with the colours corresponding to the responses in the right column.} 
    \label{fig:sedsnaps_disc}
\end{figure*}

We illustrate the model by considering 
a short X-ray flash, with amplitude $L_{x, \text{max}}/\langle L_{x} \rangle = 2$, modelled by a top hat with width $\Delta t = 0.2$\,days. We consider how this propagates down through the SED, as the flash travels across the disc. We fix the mass and inclination to that of F9, set the outer edge of the disc at $r_{\text{out}}=10^{3}$ and assume zero spin and a mass accretion rate which is 10\% of Eddington. We assume for simplicity that there is no warm Compton region, so the standard disc extends from $r_{\text{out}}$ to $r_{h}$, and take $r_{h} = 12$ and $33$ in order to see how increasing truncation of the inner disc changes both the spectrum and the response. 

\begin{figure*}
    \centering
    \includegraphics[width=\textwidth]{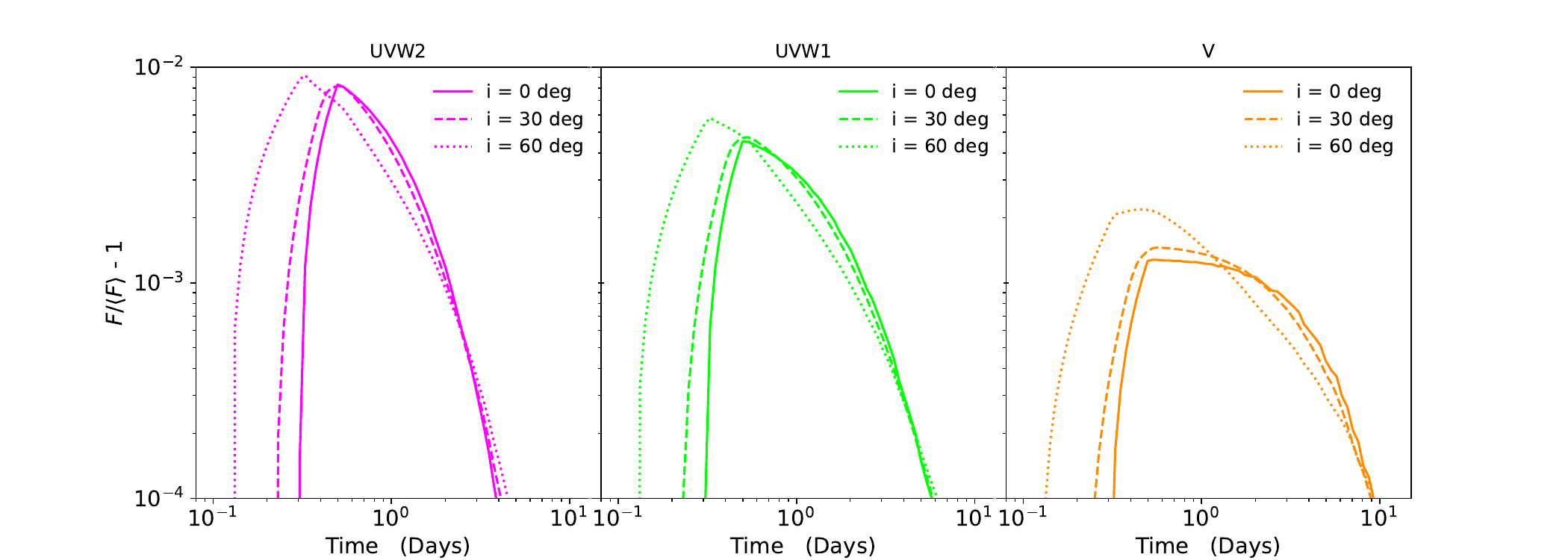}
    \caption{Response functions extracted for UVW2 (left, magenta), UVW1 (middle, lime green), and V (right, orange), as seen by an observer at inclination 0\,deg (solid), 30\,deg (dashed), and 60\,deg (dotted). These have been calculated for a disc with $r_{h}=12$.}
    \label{fig:rsp_incComp}
\end{figure*}

The left panel of Fig. \ref{fig:sedsnaps_disc} shows snapshots of the model for a face on observer. The illuminated ring propagates outwards, with times  $t = 0.5$ (blue), $1$ (orange), $2$ (green), $4$ (red) and $8$ (magenta) days after the X-ray flash.
The middle panel shows the SED (upper), with the change in spectrum at each time (lower). 
The spatial disk response is plotted on a log scale, so the constant width 0.2 light day travel time of the step function is progressively smaller on the log scale at larger radii. This then also explains why the amplitude of the fluctuation drops as the step function sweeps across the disc. Both X-ray irradiation and intrinsic flux go as $F \propto r^{-3}$, so the steady state disc temperature goes as $T\propto r^{-3/4}$ i.e. drops by a factor 2 for every factor 2.5 increase in radius. The disc can be approximated by a series of rings, each with temperature a factor 2 lower for radius increasing by a factor 2.5. 
The innermost ring is completely illuminated by the fixed width 0.2 light day flash, so it responds to the entire factor 2 increase in X-ray flux. However, at the outer radius, the light flash ring only covers a small fraction of the lowest temperature emitting region, so the amplitude of the response is much smaller. Thus the largest amplitude reverberation signal is always expected on the inner edge of the truncated disc, and the change in disk SED at all energies is dominated by the inner disc. This explains the shape of the lightcurves in UVW2 (magenta) and UVW1 (green) shown in the right hand panel. These both peak on a timescale corresponding to the light travel time to the inner edge of the disc of $[(r^2+h_{x}^2)^{1/2} +h_x]R_g/c=0.29$~days for $r_h=12$. The decay is the exponential, with a timescale roughly given by the timescale at which the illuminating flash reaches the radius with temperature which peaks in each waveband ($\sim 6$eV for UVW2 and $\sim 4$~eV for UVW1). It is this exponential decay rather than the peak response which encodes the expected
increasing timescale behaviour from a Shakura-Sunyaev disc, where $\tau\propto \lambda^{4/3}$ 
\citep{Collier99, Cackett07}, so that the decay in UVW2 is $\sim 0.7$~days, while that in UVW1 is $1.1$~days

The lower panel of Fig. \ref{fig:sedsnaps_disc} shows the
effect of increasing the disc truncation radius to $r_h=33$.
Here the light travel time to the inner edge of the disc is 0.5 days, so the blue ring showing the position of the flash on the disc after 0.5 days does not exist. Other differences are the SED (middle panel) shows a stronger hard X-ray tail, as expected as the 
higher $r_{h}$ means a larger fraction of the accretion power is dissipated in the hot Compton component. This power is taken from the inner edge of the disc, so the disc SED peaks at lower energies as well as being less luminous. 
This shifts the predicted SED peak from being in the unobservable FUV,
highlighted by the pink shaded region, for $r_h=12$, 
to emerging into the observable UV bands as shown in the lower panels of Fig. \ref{fig:sedsnaps_disc} for $r_h=33$.

The stronger hard X-ray flux for $r_h=33$ means that a factor 2 change gives a stronger response compared to the $r_h=12$ for each time segment. While the underlying disc temperature is the same, the stronger illuminating flux 
gives a higher temperature fluctuation at 1 day delay (orange line, middle lower panels, compare between $r_h=12$ and $33$).

The position of the truncated disc edge also depends on the black hole mass and accretion rate as well as $r_h$, but generally the UVW2 continuum does not sample the SED peak, even with a moderately truncated disc as assumed here. However, the lightcurve response does. The lightcurve in any wavelength band on the disc is dominated by the contribution at that wavelength from the inner edge. The lightcurve in any wavelength band has a peak response lagged by the light travel time to the inner disc edge, and then has an exponential decay whose timescale encodes the expected $\lambda^{4/3}$ dependence.

Inclination increases the light travel time smearing as expected from $\Delta \tau(r) \approx R/c(1- \sin(i))$ for the near side of the disc, to $\approx R/c(1+ \sin(i))$ for the far side. Fig. \ref{fig:rsp_incComp} compares the UVW2, UVW1 and V band response for a face on disc with that for $i=0$, $30$, and $60$\,deg, with $r_{h}=12$. As expected, we see that an increased inclination puts an additional smearing on the response function, increasing its width. Additionally it is also seen that as we increase the inclination the beginning of the response is shifted to earlier times by $\Delta \tau = (R_g/c)[h_{x}(1-\cos(i))+r\sin(i)]$ with respect to $i=0$\,deg, corresponding to $\Delta \tau \approx 0.08$, $0.17$\,days for $i=30$, $60$\,deg respectively. This is simply a reduction in the light-travel time to the inner edge of the disc on the side of the observer.

\begin{figure}
    \centering
    \includegraphics[width=\columnwidth]{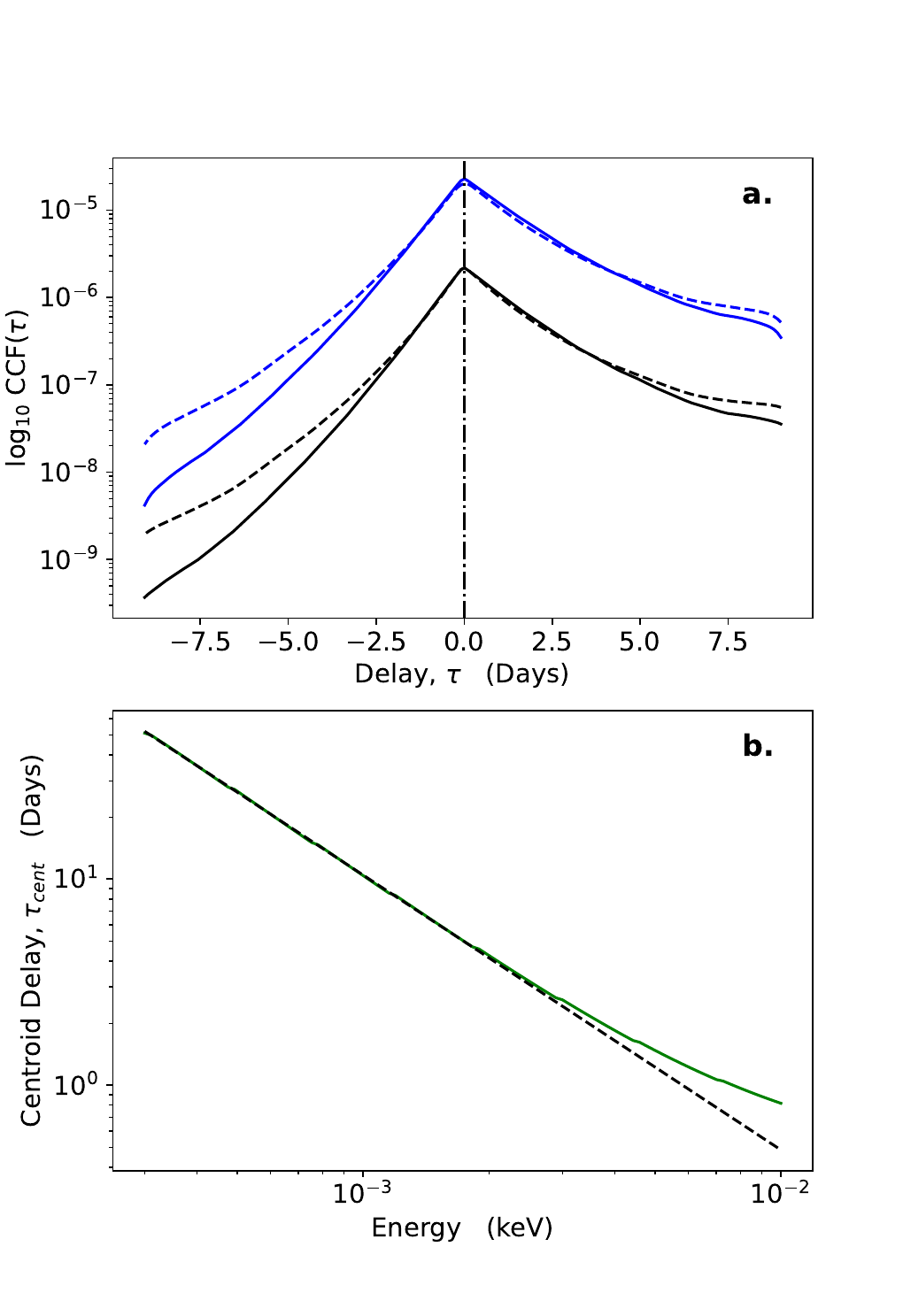}
    \caption{\textit{Top}: The cross correlation between the UVW2 and UVW1 response functions. Blue lines show the CCF for $r_{h} = 33$, while black lines show for $r_{h}=12$. Solid lines show inclination 0\,deg, while dashed show inclination 30\,deg. The dashed-dotted vertical line indicates 0 delay, where we clearly see that all the CCFs peak.
    \newline
    \textit{Bottom}: The centroid delay predicted by our model (green), and the analytic $\tau \propto \lambda^{4/3}$ relation (black, dashed line). Note, to avoid the model lag being affected by the outer edge of the disc, at low energies, we have increased the disc size to $10^{5}$ for the purpose of calculating $\tau_{\text{cent}}$.}
    \label{fig:rspCCF}
\end{figure}


Since the responses peak at the same lag for varying energy bands, we would expect their cross-correlation to peak at zero. Additionally, since they have a roughly exponential shape, we would also expect the cross-correlation to be an asymmetric function, where the asymmetry comes from the difference in decay times. This is shown in Fig. \ref{fig:rspCCF}a for a cross-correlation between the lightcurves in UVW2 and UVW1 with $r_{h}=12$ (black) and $33$ (blue), at inclinations 0 (solid) and 30\,deg (dashed). These show asymmetric functions as expected, with a reduction in the decay at higher inclination, arising from an increase in the width of the response function.

Clearly then the observed inter-band lags in AGN cannot arise from differences in the peak response time, as there is no difference in peak response time. 
Instead the increased response width at lower energies will lead to an increased mean delay, which give the observed $\tau \propto \lambda^{4/3}$ lag relation (e.g \citealt{Edelson19, Cackett20, Vincentelli21}). Fig. \ref{fig:rspCCF}b shows the mean centriod delay predicted by our response functions in solid green. These mostly follow the analytic $\tau \propto \lambda^{4/3}$ relation (black dashed line), apart from at higher energies originating close to the peak of the SED.

Our model can predict both the amount and the shape
of the response at a given energy, through explicitly considering the energetics and geometry of the system and using these to calculate a set of time-dependent SEDs. These re-produce the analytically predicted lag-energy relation when the energies considered are away from the optically thick disc peak. We now use this model to generate the lightcurves expected in any band, with smoothing caused by both the distribution of light-travel times to any given disc annulus and the continuum nature of the response. 

\section{The Data}

\subsection{The light-curves}

Fairall 9 has recently been the subject of an intensive monitoring campaign, using \textit{Swift} and Las Cumbres Observatory (LCO) \citep{Hernandez20}. We use year 1 light-curves obtained by \textit{Swift} (provided by Jaun V. Hern\'{a}ndez Santisteban and Rick Edelson through private communication). These light-curves cover the \textit{Swift}-XRT \citep{Burrows05} hard X-ray band (1.5 - 10\,keV, henceforth HX), soft X-ray band (0.3 - 1.5\,keV), and the \textit{Swift}-UVOT \citep{Roming05} broad-band filters UVW2, UVM2, UVW1, U, B, and V. They have a cadence of $\sim$1\,day, and a duration of 300\,days (from MJD 58250 to 58550). Since we expect the disc emission to peak around the UVW2 band, and since this has the cleanest data, we will limit our analysis to the HX and UVW2 light-curves. A detailed description of the data reduction is given in \citet{Hernandez20}.

The light-curves begin to rise beyond $\sim$ 58500\,MJD, which we speculate might be due to a change in the accretion structure. In order to simplify our analysis we therefore discard the section of the light-curves beyond 58500\,MJD, as a change in accretion structure would significantly complicate our results. Instead this will be the focus of a future paper. (Note, the full light-curves can be found in \citealt{Hernandez20})

In later sections we will compare light-curves using cross-correlation methods. We will also be evaluating their Fourier transforms. These techniques all rely on evenly sampled data (\citealt{Gaskell87, Uttley14}, \citetalias{Gardner17}). Since real data will not be exactly evenly sampled, we linearly interpolate the raw light-curves onto identical grids with width $\Delta t = 0.5$\,days, and then re-bin onto a grid with $\Delta t = 1$\,day.

\subsection{Extracting the spectral energy distribution}
We now extract the time-averaged SED for F9, using the spectral fitting package {\sc xspec} v.12.12.0 \citep{Arnaud96}, and model the data using {\sc agnsed} \citepalias{Kubota18}; as described in section 2. We stress that we use a slightly modified version of {\sc agnsed}, which includes heating from both sides of the disc when determining the re-processed temperature. This then gives us constraints on both the energetics and physical parameters of the system, which we will use as our base model when analysing the light-curves.

We start by extracting the time-averaged X-ray spectrum. \textit{Swift}-XRT has limited effective area for spectroscopy, but \textit{NICER} was also monitoring the source at this time. However, \textit{NICER}
has its own challenges regarding the background estimation.
Hence instead we use XMM-Newton for a detailed spectral description. The archival observation on 9 May 2014 by \citet{Lohfink16}
has soft X-ray spectrum which is compatible with \textit{NICER} and harder X-ray spectrum compatible with \textit{Swift}-XRT. It also 
has UVW1 flux from the OM within $\sim 5$\,\% of the {\it Swift}-UVOT UVW1 flux, confirming that the source was in similar state at this time.
We give more details in appendix D. 

\begin{table}
    \centering
    \begin{tabular}{c|c|c}
    \hline
    Component & Parameter (unit) & Value \\
    \hline
    {\sc phabs} & $N_{H}$ ($10^{20}$ cm$^{-2}$) & 3.5 \\
    \hline
    {\sc agnsed} & M ($M_{\odot}$) & $2 \times 10^{8}$ \\
                \\
                 & dist (Mpc) & $200$ \\
                 \\
                 & $\log(\dot{m})$ ($\dot{M}/\dot{M}_{\text{edd}}$) & $-1.215^{+0.024}_{-0.027}$ \\
                 \\
                 & $a_{*}$ & 0 \\
                 \\
                 & $\cos(i)$ & 0.9 \\
                 \\
                 & $kT_{e, h}$ (keV) & 100 \\
                 \\
                 & $kT_{e, w}$ (keV) & $0.331^{+0.042}_{-0.035}$ \\
                 \\
                 & $\Gamma_{h}$ & $1.921^{+0.026}_{-0.027}$ \\
                 \\
                 & $\Gamma_{w}$ & $2.782^{+0.030}_{-0.032}$ \\
                 \\
                 & $r_{h}$ ($R_{G}$) & $18.8^{+1.1}_{-0.9}$ \\
                 \\
                 & $r_{w}$ ($R_{G}$) & =$r_{\text{out}}$ \\
                 \\
                 & $r_{\text{out}}$ ($R_{G}$) & $2.091^{+0.075}_{-0.058}$ \\
                 \\
                 & $h_{x}$ ($R_{G}$) & 10 \\
                 \\
                 & redshift & $0.045$ \\
    \hline
    {\sc rdblur} & Index & -3 \\
                 \\
                 & $r_{\text{in}}$ ($R_{G}$) & $382^{+492}_{-160}$ \\
                 \\
                 & $r_{\text{out}}$ ($R_{G}$) & $10^{6}$ \\
                 \\
                 & Inc (deg) & 25 \\
    \hline
    {\sc pexmon} & $\Gamma$ & =$\Gamma_{h}$ \\
                 \\
                 & E$_{c}$ (keV) & $10^{4}$ \\
                 \\
                 & redshift & $0.045$ \\
                 \\
                 & Inc (deg) & 25 \\
                 \\
                 & Norm ($10^{-3}$) & $4.51^{+0.76}_{-0.68}$ \\
    \hline
    \hline
    $\chi^{2}$/d.o.f & 242.77/168 = 1.45 \\
    \hline
    \end{tabular}
    \caption{Best fit parameters for our SED model. Values with no error were frozen in the fitting process. Note that the inner radii given in {\sc agnsed} and {\sc rdblur} are not the same. This is because we only include {\sc rdblur} in order to fit the iron line profile. However, since the main focus of this paper is continuum reverberation this is only a convenience component. Hence, we will only consider the {\sc agnsed} parameters in our modelling.}
    \label{tab:SED_fitPars}
\end{table}

We also use the UV continuum in our spectral modelling, as it is in this energy range where the disc should contribute most. We start by considering the time-averaged, host galaxy subtracted flux from each UVOT filter during the campaign, given in \citet{Hernandez20}. We use the conversion factors given in Table 10 in \citet{Poole08} in order to convert to a count-rate, allowing the use of {\sc xspec} in the fitting process.

\begin{figure}
    \centering
    \includegraphics[width=\columnwidth]{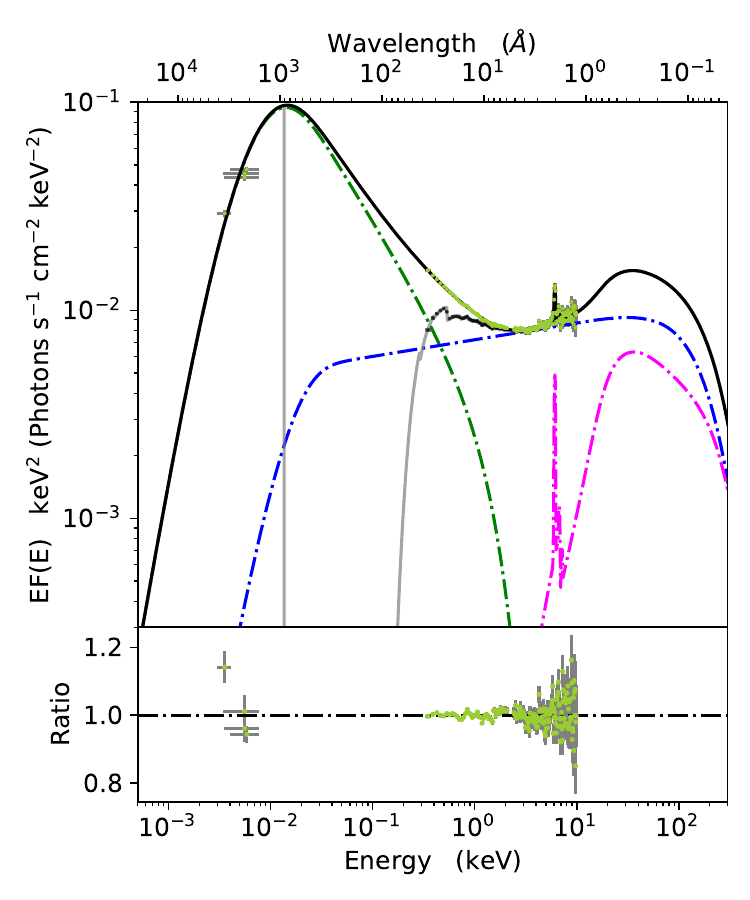}
    \caption{Time-averaged absorption corrected SED of F9. The solid black line shows the best fit spectrum, the dashed lines show the model components; which are: warm Comptonised region (green), hot Comptonised corona (blue), and neutral reflection (magenta); and the solid gray line shows the model before correcting for absorption. The residual shows the ratio between the best fit model and the data.}
    \label{fig:SED}
\end{figure}

Finally, we model the intrinsic SED following section 2, by using an updated version of {\sc agnsed} \citepalias{Kubota18}. In addition we include a reflection component, modelled with {\sc pexmon} \citep{Nandra07, Magdziarz95}, to model the Fe-K$\alpha$ line and Compton hump, along with {\sc rdblur} \citep{Fabian89} to account for any smearing in the reflection spectrum.
The detailed fits of \citep{yaqoob16} show that the iron line and reflection hump in this source are consistent with neutral material, corroborating our choice of reflection model. 
We also include a global photoelectric absorption component, {\sc phabs}, to account for galactic absorption. The final model is {\sc phabs*(agnsed + rdblur*pexmon)}. Fig. \ref{fig:SED} shows the final SED model after correcting for absorption.

While fitting the SED we freeze $kT_{e, h}$ to 100\,keV, as we do not have sufficiently high energy coverage to clearly determine the electron temperature. We also find that the data suggest strong preference to a large warm Comptonised region, leading to a negligible contribution from the standard disc region. Hence, we simply set $r_{w} = r_{\text{out}}$, as this does not alter the fit statistic, and allows us to eliminate a free parameter. We also fix the galactic absorption column at $N_{H} = 0.035 \times 10^{22}$\,cm$^{-2}$. The best fit parameters are shown in Table \ref{tab:SED_fitPars}, and the SED is shown in Fig. \ref{fig:SED}. This forms our baseline model for the following analysis.

While the soft Comptonisation gives a different spectrum from each disc annulus than the blackbody assumed in section 3, its seed photons are assumed to come from reprocessing on an underlying passive disc structure. The soft spectral index means that the warm comptonised emission peaks at an energy which tracks the seed photon energy, so it has the same $\tau\propto \lambda^{4/3}$ behaviour as a pure blackbody disc. Fig. \ref{fig:SEDsnaps_warm} in the Appendix shows the equivalent of Fig. \ref{fig:sedsnaps_disc} for this specific model. 

\begin{figure*}
    \centering
    \includegraphics[width=\textwidth]{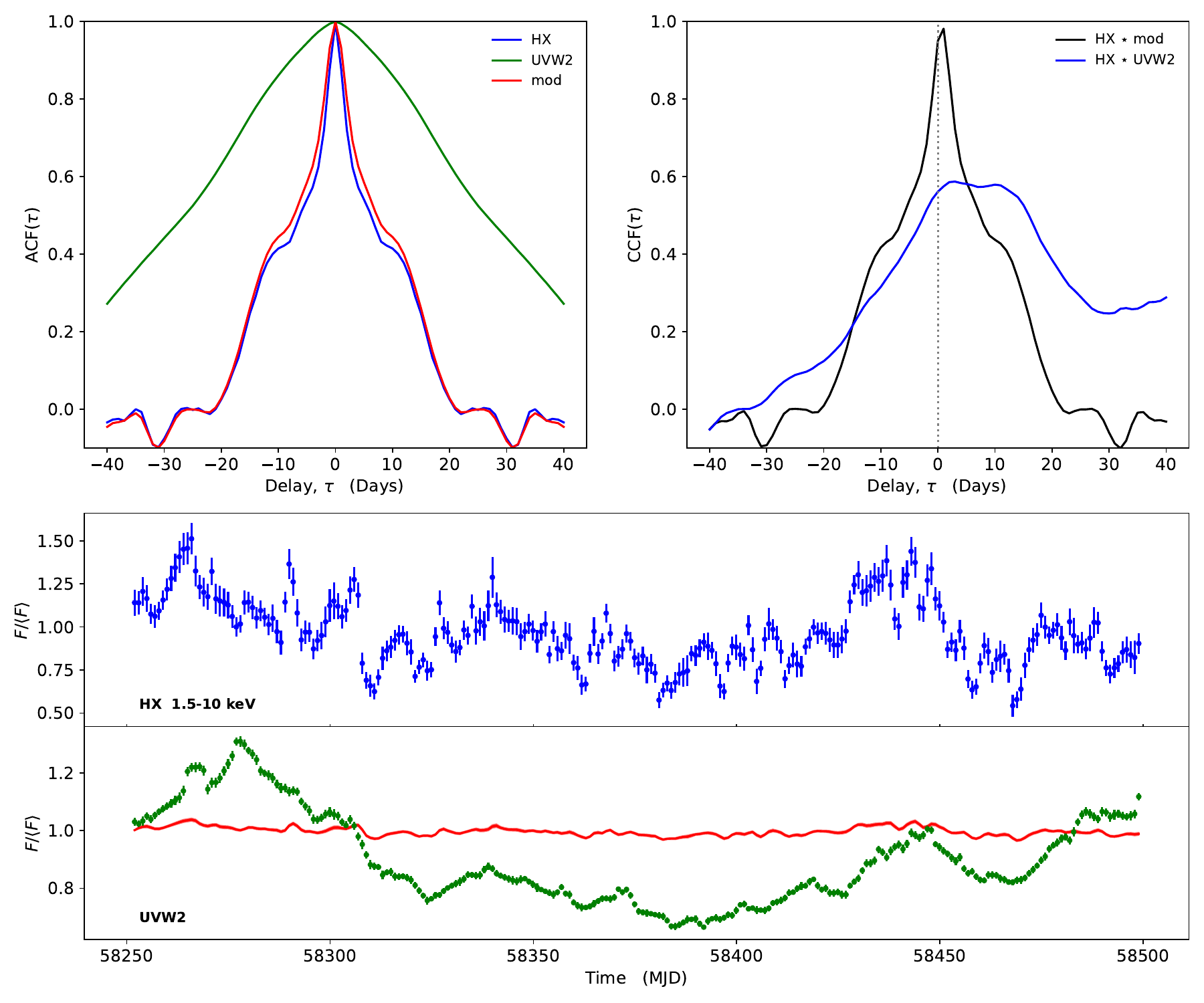}
    \caption{\textit{Top left}: The auto-correlation functions for each light-curve. The colours corresponds to the light-curves in the bottom two panel. It is clear the UVW2 is dominated by long-timescale variations, giving the broad ACF, whereas HX appears to have at least two different variability time-scales; one rapid giving the narrow peak, and one slightly slower giving the broad base. These variations are not smoothed out by disc re-processing, clearly indicated by the similarity between the model ACF and the HX ACF.
    \newline
    \textit{Top right}: The cross-correlation functions between HX and the model (black), and between HX and UVW2 (blue). The vertical grey dotted line shows $\tau = 0$\,days. 
    \newline
    \textit{Bottom panels}: The HX light-curve (blue, top panel), UVW2 light-curve (green, bottom panel) and resulting model light-curve (red, bottom panel). Evidently, the disc model completely under-predicts the response, and is in no way shape or form able to re-produce the observed UVW2 light-curve. Note that the $\pm 1\sigma$ error on the model light-curve is too narrow to be seen clearly in this plot.}
    \label{fig:modLcurve_agnsed_Raw}
\end{figure*}

\section{Comparing model and Observed Light-curves}
\subsection{Comparison to the Unfiltered Variability}

We now construct model UVW2 light-curves for F9, and compare to the observations, working on the assumption that the observed HX light-curve drives all of the variability in the UV. The variations in coronal luminosity are then the same as variations in the HX light-curve, as described in section 2.3. To generate model light-curves from the resulting time-dependent SEDs, we use the \textit{Swift}-UVOT UVW2 response matrix \citep{Roming05} to extract the part of the SED in the correct energy range and account for the energy dependent sensitivity across the filter. It is important to note that the fluxes in our driving X-ray light-curve have errors, which need to be propagated through the model. To do this we take inspiration from the flux-randomisation method, often used in determining the uncertainty on cross-correlation lags \citep{Peterson98}. For each data point in the X-ray light-curve we assign a Gaussian probability distribution, centred on the measured flux value and with a standard deviation set by the error-bar. We then then draw 5000 realisations of the X-ray light-curve, with fluxes sampled from their probability distribution, and evaluate our model light-curve for each realisation. For each time-step within the input light-curve we then have a distribution of 5000 model evaluations. Our set of model light-curve fluxes are then defined by the 50th percentile in each of these distributions, with the 16th and 84th percentiles defining the standard deviation for each model point.

Our resulting model light-curve is shown in the bottom panel of Fig. \ref{fig:modLcurve_agnsed_Raw}, along with the observed ones. Clearly the model is not consistent with observations. The amplitude of variability in UVW2 is dramatically underestimated. 

The SED we have derived for Fairall 9 is clearly disc dominated, with the UV disc component responsible for $\sim 77$\,\% of the total power, compared to the X-ray corona only contributing $\sim 23$\,\%. This tells us that when we calculate the effective temperature across the disc we are strongly dominated by the intrinsic disc emission, and so any changes in X-ray illumination will have a minimal effect on the SED. Essentially, there is not enough power in the variable X-ray lightcurve to re-produce the observed variability amplitude in the UV through re-processing alone. In fact, it is clear that the majority of the variability must be driven by some other process than re-processing of the assumed isotropic X-ray emission. The case for an alternative process is reinforced when we examine the long term trend in the observed light-curves. It is clear that the long-time scale variations in the UVW2 are not present in the X-ray. In fact, this was pointed out by \citet{Hernandez20}, who de-trended their light-curves by fitting a parabola. If all of the UVW2 variability was driven by re-processing from the X-ray corona, then we would expect to see the long term trend in the HX light-curve too. There is simply no way to create this trend purely through re-processing.

The inability of reprocessing to match observations becomes even more apparent when we examine the auto-correlation functions (ACF) of the light-curves. The HX ACF appears to contain two distinct components, one narrow due to the rapid variability, and one broad arising from longer term variations. UVW2, on the other hand, has a much broader ACF, indicating that the majority of the UVW2 variability comes from longer-term fluctuation than seen even in the longest timescale in the hard X-rays. If disc re-processing was the sole driver of the UVW2 variability, then we would expect our model ACF to match the UVW2 ACF. Clearly this is not the case. The model ACF is almost identical to that of the HX ACF, albeit with a slight broadening at the bottom of the narrow component. If disc re-processing was giving a significant contribution to the total UVW2 variability we could expect that the model ACF should at least be somewhere in between that of HX and UVW2. Again, this is not the case.  

To really highlight the lack of  impact of reprocessing in making the UVW2 lightcurve we show the cross correlation functions (CCF). The HX and UVW2 are poorly correlated. More interestingly, perhaps, is the correlation between our model and the HX light-curve. The similarity between this CCF and their respective ACFs suggest that the time-scales over which the model light-curve is smeared are far too small to make a significant impact. This is unsurprising when we examine the width of the narrow component in the CCF, and their ACFs, which we expect would be the first thing that would be smoothed out. We already know, from section 2.4, that the majority of the response will come from the inner edge of the disc, which for the SED fit derived in section 3 is at roughly $\sim 5-6$\,light-hours. This is considerably shorter than the width of the narrow component in the CCF/ACFs, which is on the time scale of a few days. Although we also expect an increase in smoothing due to the continuum nature of the response, it is not expected that this would increase the time-scale sufficiently to wipe out the rapid variations. In other words, the rapid variability seen in the X-ray light-curve is on time-scales longer than the smoothing imposed by the re-processing model. Hence, this rapid variability must also be present in our model light-curve, which explains the near identical nature of the model and HX ACFs, along with the strong similarity between these ACFs and their CCF. 

These results clearly show that AGN continuum variability is somewhat more complex than can be described by re-processor models. However that is not to say that re-processing does not take place, just that it cannot be the dominant source of variability. Hence, if we wish to study the continuum re-processor, we need a way of extracting and disentangling the different sources of variations from observed light-curves; the focus of the next subsection.

\subsection{Isolating and Exploring the Short-Term Variability}

When trying to isolate the re-processed variability within the data, we first need some idea of what time-scales we expect the re-processing to occur over. The clear choice, from examining the light-curves, is to filter out the long term variations in the UVW2, as these are not present in either the HX light-curve or its ACF. If all variability was driven by re-processing of X-rays, then we would expect to see the large trends within the UVW2 light-curve in the HX one too. However, that is clearly not the case, as we see no indication of the long $\sim 100$\,day trend, which is so clearly present in UVW2, in HX. This also fits nicely into our issue with the energetics not producing enough response. The UVW2 variability is clearly dominated by this trend, and so filtering it out should reduce the variability amplitude.

Previous studies have removed long term trends by fitting a function that roughly matches the observed shape of the trend-line, e.g. the parabola of \cite{Hernandez20}. This is useful in the sense that it provides a simple method for estimating the long-term time-scales involved, and for recovering the shape of the short-term variability. However, it dependent on the function chosen fit to the trend, and does not give an insight into the variability power produced on different time-scales. The latter point is very important, as we have seen in the previous sections that the system energetics determine the amplitude of the re-processed variability. So in order to constrain both the variable power and variability time-scales within the data we use a Fourier based approach to de-trend our light-curves.

\begin{figure}
    \centering
    \includegraphics[width=\columnwidth]{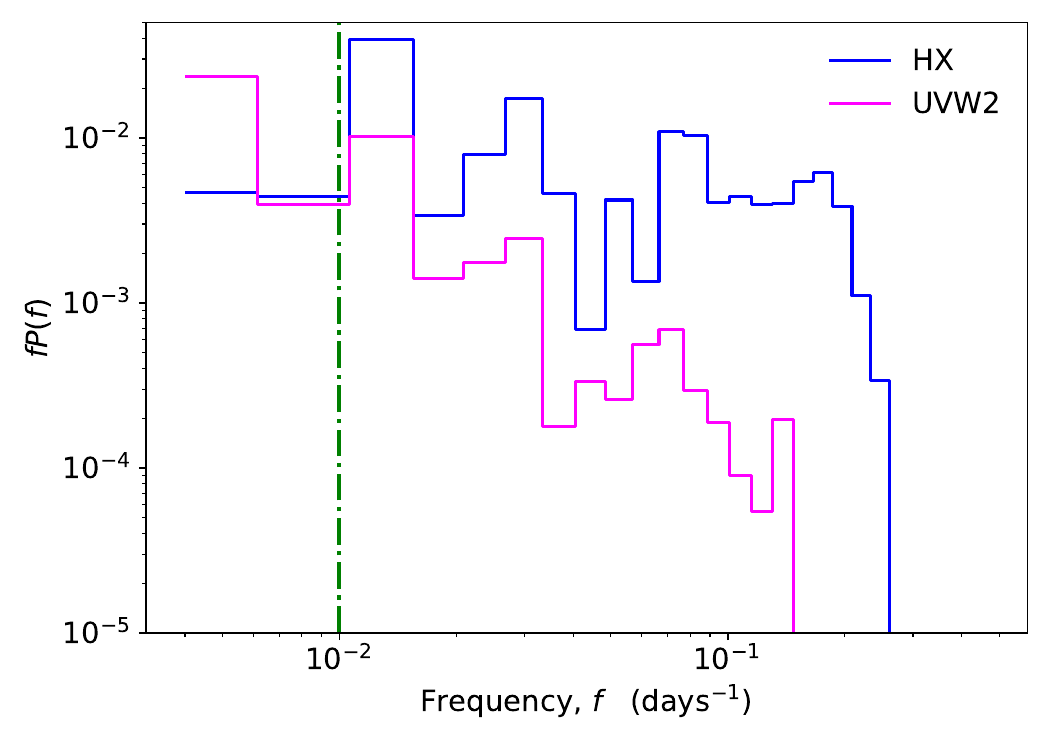}
    \caption{Power spectrum of HX (blue) and UVW2 (magenta) light-curves. These have been re-binned, with $d\log f = 0.1$ days$^{-1}$. The green dashed line indicates the frequency cut, below which we set the UVW2 power to 0 when performing the Fourier filtering. Note that due the limited number of points in the light-curve we have not averaged over multiple segments; as this would remove the long term trend. Hence the errors on the PSD will be equal to the power within each frequency bin. We have chosen not to plot the error bars, as these would make the figure incredibly unclear. Also, we highlight that this PSD is only used as a guide for the filtering process.}
    \label{fig:raw_powerSpec}
\end{figure}

\begin{figure*}
    \centering
    \includegraphics[width=\textwidth]{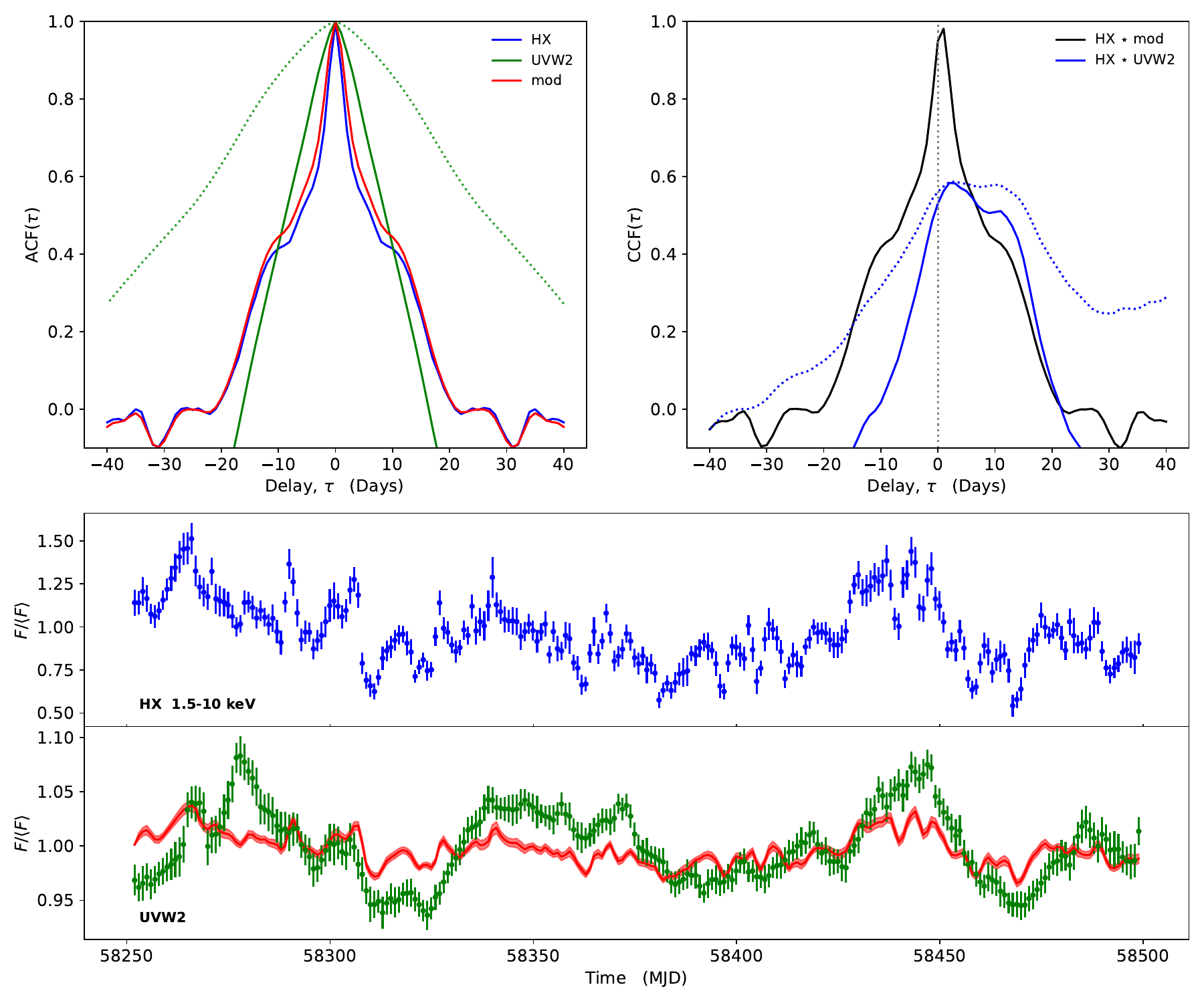}
    \caption{Same as Fig. \ref{fig:modLcurve_agnsed_Raw}, but now using the filtered UVW2 light-curve. The dotted lines in the top left and right panels show the ACF and CCF, respectively, for the unfiltered UVW2 light-curves in order to highlight the effect the filtering process has on the correlation functions.}
    \label{fig:modLcurve_agnsed_Filt}
\end{figure*}

Extracting information about variability on different time-scales through Fourier analysis is already commonly done in accretion studies, mostly for rapidly varying objects such as accreting white dwarfs, neutron stars, or black hole binaries (see \citealt{Uttley14} for a review); but are also increasingly common in AGN studies (e.g \citealt{McHardy05, McHardy06, Kelly11}). Previous studies focus on modelling the power spectral density (PSD), and use this to understand what processes are occurring on different time-scales. Instead, we will simply use the PSD to estimate the time-scales we need to filter out of our light-curves.

We start by calculating the PSD for both the UVW2 and HX light-curves, using the {\sc python} spectral timing analysis package {\sc stingray}\footnote{DOI: 10.5281/zenodo.6290078} \citep{Huppenkothen19}. Unlike the black hole binaries, our lightcurve is not long enough to 
calculate the averaged PSD from multiple segments, so we increase the samples at each frequency by geometric re-binning of $d\log f = 0.1$\,days$^{-1}$. 
This is shown in Fig. \ref{fig:raw_powerSpec}.
Nontheless,  the 
PSD at the lowest frequencies/longest timescales consist of only a few (or even single) point, where the uncertainty on the power is equal to the power, but the trend is clear. There is more power in the UVW2 variability at low frequencies, $f \lesssim 0.01$\,days$^{-1}$, than in the HX. This shows that these long timescale variations cannot be driven by reprocessing. There is no clear point at which the UVW2 power spectra change, so we place a cut-off frequency at $f_{\text{cut}} = 0.01$\,days$^{-1}$, below which we consider all UV variability to belong to the long term trend. This will probably be an underestimate of the intrinsic UV variability.

We filter the UVW2 lightcurve on this frequency by taking its discrete Fourier transform using the {\sc scipy.fft} library \citep{Virtanen20} in {\sc python}. Discarding everything below $f_{\text{cut}}$, and taking the inverse Fourier transform back again, gives a light-curve only displaying variability on time scales shorter than 100 days. The bottom panel in Fig. \ref{fig:modLcurve_agnsed_Filt} shows the resulting UVW2 light-curve in green. As designed, the long-term trend that was previously present has been completely removed, while preserving the short term variations. Additionally, the variability amplitude has been reduced, which is as expected considering the variability power is dominated by the low-frequency variations. We stress that we have only filtered the UVW2 light-curve. Since our model assumes all the X-rays originate from the central region, then if re-processing occurs it should include all the X-ray variability. Hence, the HX light-curve should remain un-filtered.

We now perform the same analysis as in section 4.1, and show the results in Fig. \ref{fig:modLcurve_agnsed_Filt}. Immediately we see that, although closer, our model light-curve still does not match the data. The variability amplitude for the fast fluctuations is still underestimated, albeit by a much smaller factor than previously. In fact,  by eye, there are clear similarities, particularly the dip at $\sim 58310$\,MJD, and the rise $\sim 58430$\,MJD. Increasing the response though would not solve some of the underlying issues which is that our model predicts a much faster UVW2 response to the HX fluctuations. 
We highlight this in the ACF, where it is clear that removing the long term UVW2 trend by filtering leads to a significant narrowing of the ACF. However, it is still not as narrow as the narrow core seen in the HX ACF 
which is imprinted onto the model UVW2 ACFs. Additionally, we can see in the CCFs that filtering has not led to a better correlation between the HX and UVW2 light-curves. The CCFs are narrower, a result of removing all long term trends, but the maximum correlation co-efficient is still roughly the same, at just less than $\sim 0.6$. This tells us that the main driver in the poor correlation between HX and UVW2 is the presence of the strong rapid variability in the X-ray, which is not at all present in UVW2. The model needs to include additional smoothing occurring on time-scales greater than the most rapid HX variability.

\subsection{Exploring the possibility for an additional re-processor}

\begin{figure}
    \centering
    \includegraphics[width=\columnwidth]{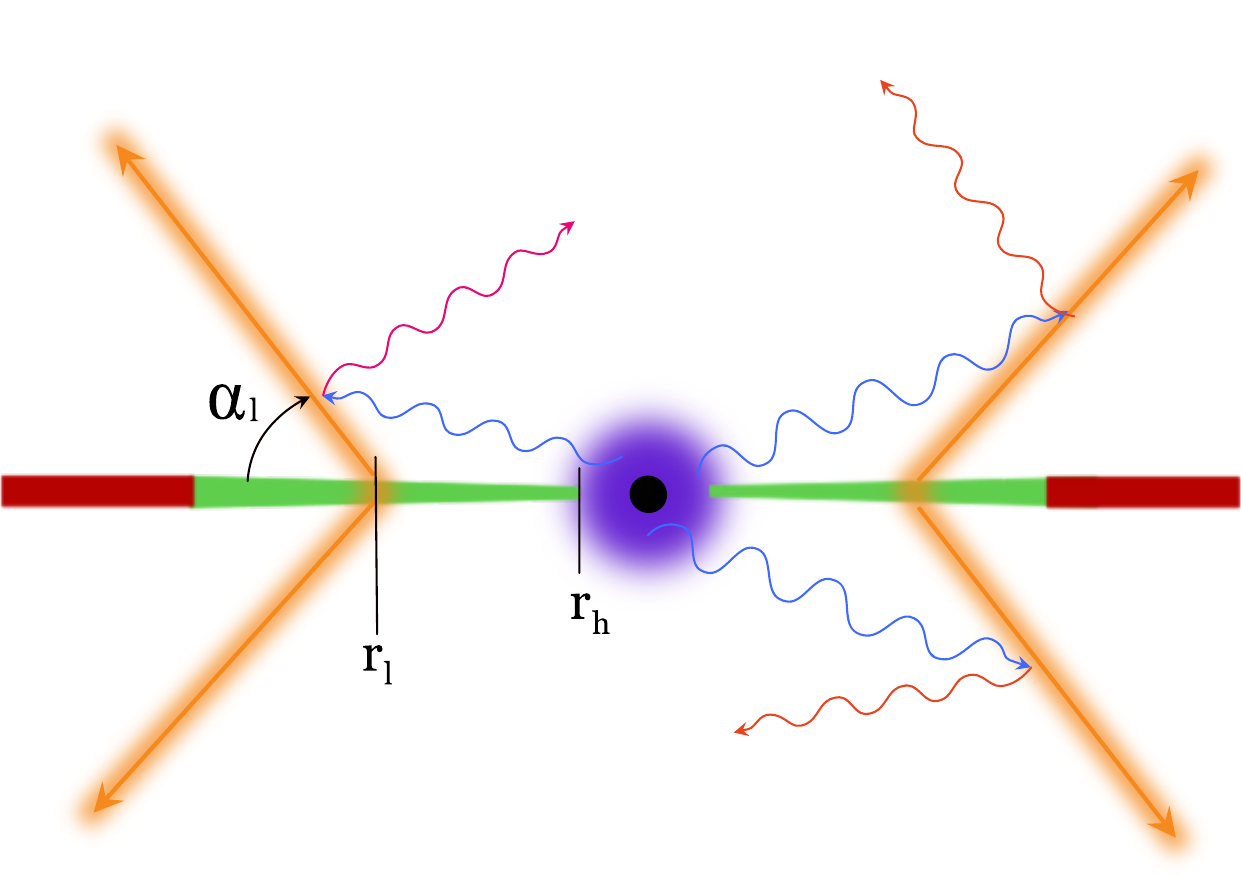}
    \caption{Sketch of the Bi-Conical geometry used for our additional re-processor. The disc and coronal regions are identical to {\sc agnsed}, while the Bi-Cone (in orange) illustrates the outflow. We assume the X-rays (blue) induce both reflected (magenta) and re-processed (orange) emission off the wind}
    \label{fig:agnref_geom}
\end{figure}

Similar issues in the previous multiwavelength campaigns have highlighted an additional contribution from broad line region (BLR) scales. The BLR must have substantial scale height in order to intercept enough UV flux to produce the line luminosities observed. This structure has a larger scale height than the flat disc, so 
has more solid angle as seen from the X-ray source, so can be an important contributor to reprocessing. 
Additionally an increased distance from the X-ray source will lead to an increased smoothing effect, potentially providing the mechanism we need to remove the fast-variability. However, increasing the distance also leads to longer lags, and we can see from the light-curves in Fig. \ref{fig:modLcurve_agnsed_Filt} that much longer lags may not work with the data either. 

The argument for reverberation from a diffuse continuum produced by the BLR or an inner disc wind is that it gives a mechanism for the longer lags seen in the data. 
This is most clearly established in the U-band, where the Balmer jump ($\sim 3600$\,\AA, commonly associated with the BLR) gives sufficiently strong emission such that the observed lag is dominated by the diffuse emission \citep{Korista01, Korista19, Lawther18, Cackett18}, but will naturally also extend to other wavebands. More recent analysis have also given results consistent with reverberation off the BLR/inner wind \citep{Dehghanian19b,Netzer22, Vincentelli22}. Thus we expect that the reprocessing structures are more complex. The disc is closest, so responds first, but then there is diffuse continuum response from the wind/BLR on longer time-scales. 

To test this we construct a simple model, containing the same disc structure as {\sc agnsed}, but with the addition of some outflow launched from radius $r_{l}$ at an angle $\alpha_{l}$ with respect to the disc. We assume this outflow contributes both to the diffuse and reflected emission, as sketched in Fig. \ref{fig:agnref_geom}.
The issue now is that we do not have a clear model of the expected emission from this component. Instead, we approximate this diffuse emission as a blackbody. Our model then, consists of the intrinsic emission dissipated in the accretion flow, calculated as in {\sc agnsed} \citepalias{Kubota18} (see section 2), followed by a black-body component located in the UV, used to approximate the diffuse contribution to the SED, and a reflected component, modelled with {\sc pexmon} \citep{Nandra07}. We assume that the
luminosity of both the reflected and diffuse emission is set by the fraction of the X-ray luminosity absorbed/reflected by the outflow. This is simply set by the covering fraction, $f_{\text{cov}}$, of the outflow as seen by the X-ray corona, and the wind albedo, $A_{w}$, such that:

\begin{equation}
    L_{\text{diff}} = \frac{1}{2} L_{x} f_{\text{cov}} (1 - A_{w})
    \label{eqn:Ldiff}
\end{equation}

\begin{equation}
    L_{\text{ref}} = \frac{1}{2} L_{x} f_{\text{cov}} A_{w}
    \label{eqn:Lref}
\end{equation}

where $L_{\text{diff}}$ is the diffuse luminosity, and $L_{\text{ref}}$ is the reflected luminosity. The covering fraction is related to the solid angle of the outflow, as seen by the X-ray corona, through $f_{\text{cov}} = \Omega/(4\pi)$. The factor $1/2$ comes from the fact that the outflow should be launched from both sides of the disc, hence the covering fraction is the total fraction of the sky covered by the outflow (this is illustrated in Fig. \ref{fig:wind_solid} in Appendix A). However, the disc will block the emission from the outflow on the underside of the disc, resulting in the observer only seeing the top-side emission; i.e half the emission. The SED model can then be described in {\sc xspec} as {\sc agnsed + $N_{\text{diff}}$*bbody + $N_{\text{ref}}$*rdblur*pexmon} where $N_{\text{diff}}$ and $N_{\text{ref}}$ are normalisation constants set to satisfy Eqn. \ref{eqn:Ldiff} and \ref{eqn:Lref}. We have also included {\sc rdblur} to account for any smearing within the Fe-K$\alpha$ line originating in the reflected spectrum. For simplicity we will refer to this model as {\sc agnref}, and make it publicly available as an {\sc xspec} model.  \footnote{\url{https://github.com/scotthgn/AGNREF}}. 

We use the covering fraction of the wind to constrain $L_{\text{diff}}$ and $L_{\text{ref}}$ in our SED model, however this does not set the absolute size scale for the wind. Hence, the SED is unable to constrain $r_{l}$ and $\alpha_{l}$. Instead we treat these as free parameters and marginalise over them in the timing analysis. Additionally, the assumed temperature of the wind will also play a role in the model, determining the position of the black-body component in the SED and affecting the variability amplitude (see Appendix B).

As in section 2.3, we calculate our model light-curve by varying the input X-ray luminosity according to an observed light-curve and creating a series of SEDs based on what X-ray luminosity each grid point in our model geometry sees at any given time. For details on how we do this for our outflow geometry, see Appendix A. However, unlike previous sections, we let the light-curves play a role in determining the time-averaged SED to determine the thermal component temperature, $kT_{\text{out}}$ as this cannot be reliably constrained through SED fits as the blackbody shape is only an approximation to the full diffuse emission. 

\begin{figure*}
    \centering
    \includegraphics[width=\textwidth]{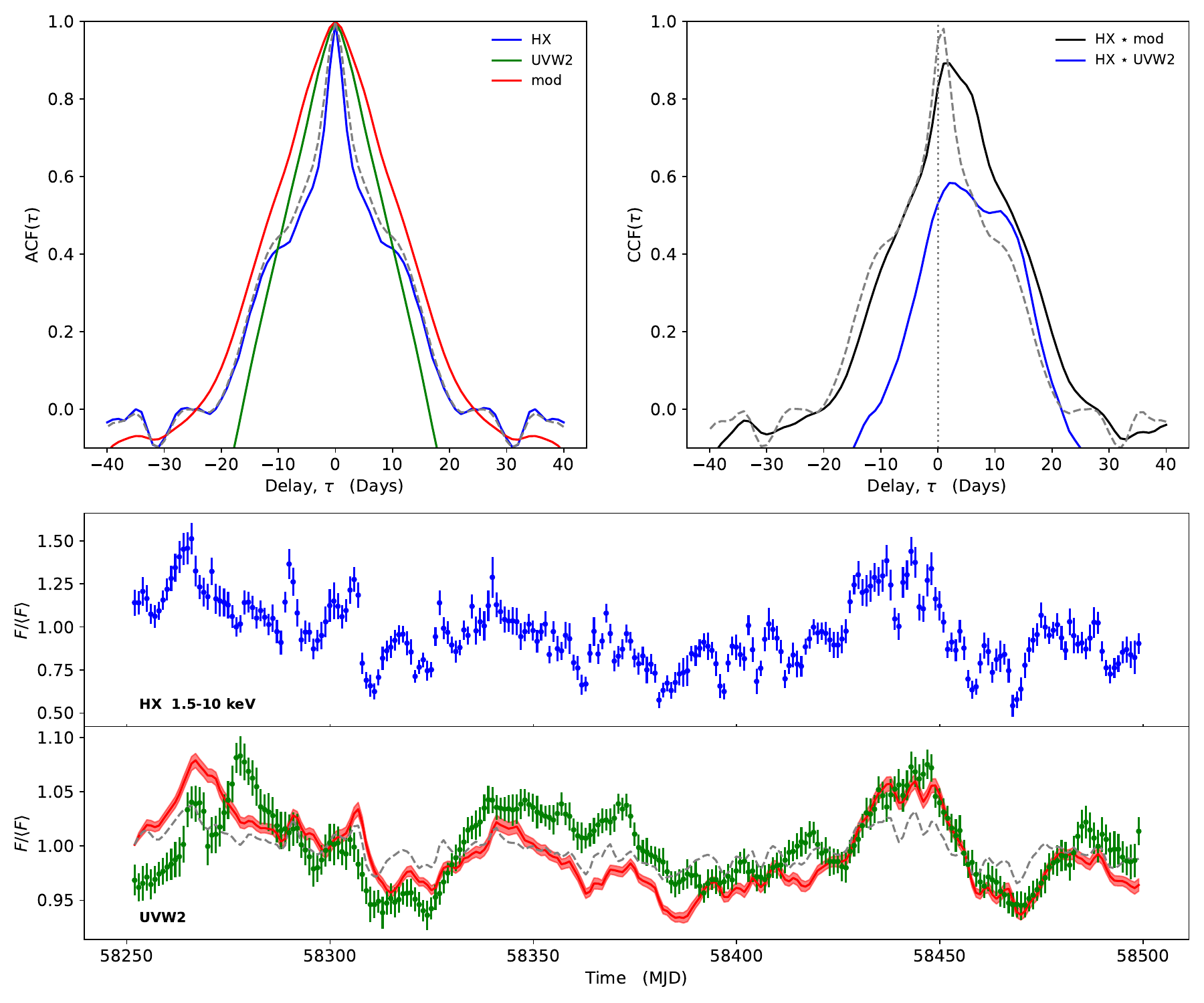}
    \caption{As in Fig. \ref{fig:modLcurve_agnsed_Raw} and \ref{fig:modLcurve_agnsed_Filt}, but for the {\sc agnref} model. The dashed grey lines show the results from the {\sc agnsed} model, in order to highlight the changes induced by including the additional re-processor.}
    \label{fig:modLcurve_agnref}
\end{figure*}

We start by defining an upper and lower limit on $kT_{\text{out}}$. As we have \textit{Swift} light-curve data extending below UVW2, down to the V band, we do this by performing a grid based search in $kT_{\text{out}}$ and comparing to all available light-curve data. We set $\Delta kT_{\text{out}} = 1\times10^{-3}$\,keV and search from $kT_{\text{out}} = 1\times10^{-3}$\,keV to $kT_{\text{out}}=1\times10^{-2}$\,keV. From Appendix B we know that $kT_{\text{out}}$ will only affect the variability amplitude of the model light-curves, and not the timing properties. Hence, for each point in our $kT_{\text{out}}$ grid we fit the model SED, calculate the corresponding model light-curves, and compare the variability amplitude to all observed bands, and check if it is over- or under-estimated. This leads to a lower limit $kT_{\text{out}, \text{low}}=2\times10^{-3}$\,keV and an upper limit of $kT_{\text{out}, \text{up}}=3\times10^{-3}$\,keV. We now refine our grid to $\Delta kT_{\text{out}}=1\times10^{-4}$\,keV and search between our lower and upper limit. However, we now also construct grids in $r_{l}$ and $\alpha_{l}$, in order to attempt to constrain the geometry of the system. To determine the radial limits we take inspiration from the HX to UVW2 CCF in Fig. \ref{fig:modLcurve_agnsed_Filt}. Although the correlation is poor, the CCF peak tentatively suggests a lag of no more than $\sim 2-4$\,days. Incidentally, this is also suggested by the ACFs, as $2-4$\,days would provide the smoothing necessary to remove the narrow core in the HX and model ACFs. Hence, we set our launch radial grid to be $200 \leq r_{l} \leq 400$ (i.e $\sim 2.2$\,light-days and $\sim 4.5$\,light-days), with a spacing of $\Delta r_{l}=10$. The lower grid limit in $\alpha_{l}$ is set by the minimum angle that still allows the outflow to obtain the required covering fraction derived in the SED fit for each value of $kT_{\text{out}}$, which is $\alpha_{l, \text{low}} \approx 65$\,deg. The maximum is simply $\alpha_{l, \text{up}}=90$\,deg, which would give a cylindrical geometry to the outflow. We set the grid spacing in $\alpha_{l}$ to $\Delta \alpha_{l} = 2.5$\,deg. This provides us with a 3D-grid in parameter space, over which we perform a parameter scan, providing 2541 potential light-curves. For each of these we calculate the CCF between the model and the UVW2 light-curve, and naively let our best fit parameters be those which give the best correlation with UVW2. This provides $kT_{\text{out}}=2.5\times10^{-3}$\,keV, $r_{l}=400$, and $\alpha_{l}=65$\,deg. The light-curve is shown in Fig. \ref{fig:modLcurve_agnref}. 

It is worth pointing out that both $r_{l}$ and $\alpha_{l}$ are at their respective grid boundaries, indicating that either we have not let our grids extend far enough or the model does not match the light-curve and so is unable to constrain the parameters. The latter seems more likely as  Fig. \ref{fig:modLcurve_agnref} shows that there is still a clear mismatch to the observed UVW2 light-curve, especially in the first half of the data.

\begin{table}
    \centering
    \begin{tabular}{c|c|c}
        \hline
        Component & Parameter (Unit) & Value \\
        \hline
        \large{{\sc phabs}} & $N_{H}$ ($10^{20}\,$cm${-2}$) & 3.5 \\
        \hline
        \large{{\sc agnref}} \\
        {\sc agnsed} & $M$ ($M_{\odot}$) & $2\times10^{8}$ \\
            \\
            & Distance (Mpc) & 200 \\
            \\
            & $\log (\dot{m})$ ($\dot{M}/\dot{M}_{\text{edd}})$ & $-1.159^{+0.027}_{-0.031}$ \\
            \\
            & $a_{\star}$ & 0 \\
            \\
            & $\cos(i)$ & 0.9 \\
            \\
            & $kT_{e, h}$ (keV) & 100 \\
            \\
            & $kT_{e, w}$ (keV) & $0.347^{+0.045}_{-0.039}$ \\
            \\
            & $\Gamma_{h}$ & $1.918^{+0.026}_{-0.027}$ \\
            \\
            & $\Gamma_{w}$ & $2.781^{+0.032}_{-0.036}$ \\
            \\
            & $r_{h}$ ($R_{G}$) & $18.1^{+1.2}_{-0.9}$ \\
            \\
            & $r_{w}$ ($R_{G}$) & =$r_{\text{out}}$ \\
            \\
            & $\log(r_{\text{out}})$ ($R_{G}$) & $1.940^{+0.064}_{-0.051}$ \\
            \\
            & $h_{x}$ ($R_{G}$) & 10 \\
            \\
            \\
            & $f_{\text{cov}}$ ($\Omega/4\pi$) & $0.852^{+0.081}_{-0.080}$ \\
            \\
            & $A_{w}$ & 0.5 \\
            \\
            \\
        {\sc bbody} & $kT_{\text{out}}$ (keV) & $2.5\times10^{-3}$ \\
            \\
            \\
        {\sc rdblur} & \textcolor{red}{Index} & -3 \\
            \\
            & $r_{\text{in}}$ ($R_{G}$) & $393^{+568}_{-167}$ \\
            \\
            & \textcolor{red}{$r_{\text{out}}$ ($R_{G}$)} & $10^{6}$ \\
            \\
            \\
        {\sc pexmon} & \textcolor{red}{$\Gamma$} & $=\Gamma_{h}$ \\
            \\
            & \textcolor{red}{$E_{c}$ (keV)} & $10^{5}$ \\
            \\
            \\
            & Redshift & 0.045 \\
        \hline
        \hline
        $\chi^{2}/\text{d.o.f}$ & 265.43/166 = 1.60 \\
        \hline
        
    \end{tabular}
    \caption{Best fit parameters for the {\sc agnref} SED model. Parameters with no errors were kept frozen during the fitting process, while parameters highlighted in red are hardwired into the {\sc agnref} model code; so are only included here for completeness. Similar to the fit in table \ref{tab:SED_fitPars}, $r_{\text{in}}$ for {\sc rdblur} is only used to set the width of the iron line, and carries no other physical meaning in our model. We have also hardwired all abundances in {\sc pexmon} to solar values.}
    \label{tab:agnref_pars}
\end{table}

Table \ref{tab:agnref_pars} give the SED parameters resulting from the above analysis. It is interesting to note how this model deviates from our original {\sc agnsed} fit. Firstly, to compensate for the additional thermal component the inner disc radius is moved slightly inwards. The model also derives a large value for the covering fraction, $f_{\text{cov}} \sim 0.87$, so that most of the reprocessing is from the wind/BLR rather than from the disc. This is most likely driven by the X-ray data, since the magnitude of iron line must be satisfied by the reflected component; while the UV points can easily compensate for changes in the thermal component by adjusting mass accretion rate and outer disc radius. The implications of this is that the diffuse reprocessed component, whose power is most probably set by the fit to the reflected component, becomes a significant factor in the SED. In terms of the light-curves this would imply that a significant portion of the variability will originate from the outflow. 

The significance of the outflow re-processed variability in the model light-curves becomes exceptionally clear in the ACF, seen in Fig. \ref{fig:modLcurve_agnref}. Here we see that the model ACF no longer resembles that of HX, unlike the case where we only considered disc re-processing. In fact, the majority of the fast variability has been completely removed, an effect induced by the smoothing. The complete removal of the narrow component in the ACF would also indicate that the outflow might even dominate the variability, unsurprisingly since the outflow will see considerably more of the X-ray power than the disc, due to the larger solid angle. 

However, we note that there is still some variability which is not captured by the model (especially before 58425\,MJD), 
likely due to both the simplistic nature of the outflow model and some further intrinsic UVW2 variability which was not filtered out by our simple Fourier filter approach. We also note that this model
gives a significantly worse fit to the SED than the original {\sc agnsed} fit. We interpret this as being due to our approximation that the diffuse emission can be modelled as a blackbody. We will use more realistic models for this in future work.

\section{Conclusions}

We have developed a full spectral timing code to calculate model light-curves in any band given a mean SED and X-ray light-curve. Our approach assumes that the X-rays are from an isotropic central source, and that there are no extrinsic sources of variability (e.g. absorption events or changing source/disc geometry). Our approach predicts the amplitude and shape of the lightcurves, not just a mean lag time, allowing a point by point comparison to the observed lightcurves. 

We apply our model to intensive multiwavelength campain dataset on Fairall 9. We fit the SED with a warm Comptonised accretion disc, plus hard X-rays produced by Comptonisation from hot plasma heated by the remaining gravitational power from within $r_h\sim20$. The predicted UVW2 
light-curve is entirely inconsistent with the observations, completely underpredicting the observed amplitude of UVW2 variability (see Fig. \ref{fig:modLcurve_agnsed_Raw}). Reprocessing of the observed hard X-ray emission cannot be the origin of the UVW2 variability in this geometry.

Nonetheless, there are features in the UV lightcurve which look like the X-ray lightcurve, so we try to isolate the  contribution of reprocessing by filtering the UVW2 lightcurve to remove the long timescale variability. This gives a closer match to our model light-curve, however the amplitude is still under predicted, and our model still varies too fast in comparison to the data. 

Recent progress in understanding the observed lags in AGN have focussed on an additional reprocessor from the BLR/inner disc wind. We approximate this emission as a blackbody and fit to the observed lightcurves to constrain the contribution of the diffuse emission from this reprocessor in the UVW2 waveband. This gives a better match to the observed variability amplitude, but it is clear that the simple filtering process did not remove all the intrinsic UVW2 variability, and that a blackbody is not a good description of the diffuse reprocessed emission. 

The intensive multiwavelength reverberation campaigns were designed to measure the size scale of the disc in AGN. Instead, the reverberation signal in the UV mostly comes from the BLR/inner wind. These campaigns also assumed that X-ray reprocessing was the main driver of variability in the UV lightcurves. This is clearly not the case in F9, as can be seen on simple energetic arguments. Unlike the lower Eddington fraction AGN (NGC5548 and NGC4151), the variable 
hard X-ray power is not sufficient to drive the variable UV luminosity. Instead, most of the variability in the UV comes from some intrinsic process in the disc itself (see also \citealt{Mahmoud22} for a similar conclusion in the similar Eddington fraction source Akn120). 
This highlights the disconnect between the timescales predicted by the standard \citet{Shakura73} disc models, where intrinsic mass accretion rate fluctuations can only propagate on the viscous timescale which is extremely long. Instead, this could favour some of the newer radiation magnetohydrodynamic simulations, which indicate a disc structure that is quite unlike these classic models, where there is considerable variability on the sound speed \citep{Jiang20}.

\section*{Acknowledgements}

We would like to thank Juan Hern\'{a}ndez Santisteban and Rick Edelson for providing the excellent data on Fairall 9. We would also like to thank the anonymous referee for helpful comments regarding the manuscript. This work made use of data supplied by the UK Swift Science Data Centre at the University of Leicester. SH acknowledges support from the Science and Technology Facilities Council (STFC) through the studentship grants ST/S505365/1 and ST/V506643/1.
C.D. acknowledges the Science and Technology Facilities Council (STFC) through grant ST/T000244/1 for support. Part of this work used the DiRAC@Durham facility managed by the Institute for Computational Cosmology on behalf of the STFC DiRAC HPC Facility (www.dirac.ac.uk). The equipment was funded by BEIS capital funding via STFC capital grants ST/P002293/1, ST/R002371/1 and ST/S002502/1, Durham University and STFC operations grant ST/R000832/1. DiRAC is part of the National e-Infrastructure

\section*{Data Availability}
The \textit{Swift} data used in this paper were provided by Jaun V. Hern\'{a}ndez Santisteban and Rick Edelson through private communication. These are described in detail in their paper \citet{Hernandez20}, and are available through the \textit{Swift} archive \url{https://www.swift.ac.uk/swift_live/index.php}. The XMM-Newton data used in the SED fits are archival, and can be directly accessed through HEASARCH (\url{https://heasarc.gsfc.nasa.gov/db-perl/W3Browse/w3browse.pl})



\bibliographystyle{mnras}
\bibliography{Refs} 




\appendix

\section{Deriving the Bi-Conical Geometry and Variability}

To understand how our model outflow responds to changes in X-ray illumination we need to be able to describe the overall geometry of the wind as seen by the BH, such that we can calculate the time delay, and how much X-ray flux each grid point sees, such that our normalisation is correct.

Our bi-conical model takes three parameter inputs to describe the global geometry: the covering fraction $f_{\text{cov}}$, the launch angle $\alpha_{l}$, and the launch radius $r_{l}$. To determine the wind variability, we need to first define a grid across the wind surface and secondly determine the time-delay at each grid point on this surface. We will start by defining our grid in terms of the polar angles $\cos(\theta)$ and $\phi$, as these can be easily related to the solid angle of each grid point, which in turn tells us the X-ray power seen by the grid point. Throughout we place the BH in the centre of our coordinate system.

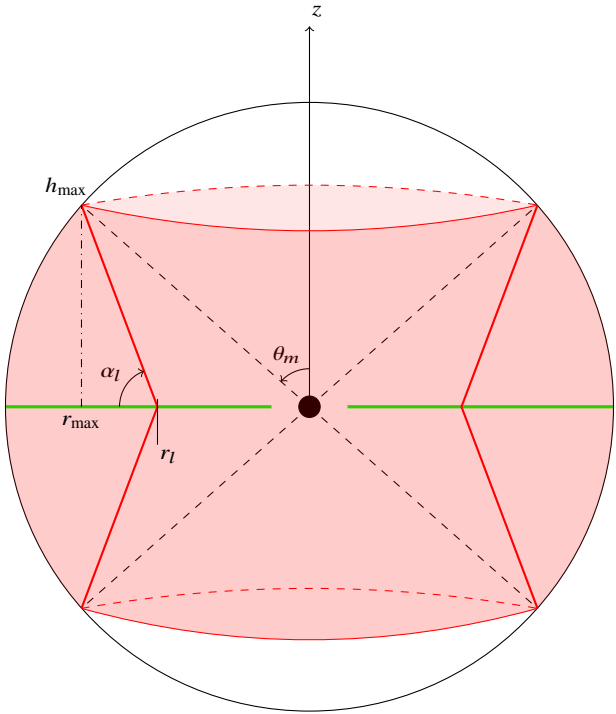
\begin{figure}
    \centering
    \begin{tikzpicture}
        
        \filldraw[black] (0, 0) coordinate (BH) circle (4pt) ;
        \draw[very thick, green] (-4, 0) coordinate(D_out) -- (-0.5, 0) ;
        \draw[very thick, green] (4, 0)--(0.5, 0) ;
        \draw[black, name path=Circ] (0,0) circle (4) ;
        \draw[->] (0, 0) -- (0, 5) coordinate (Z) ;
        \node (z) at (0.1, 5.2) {$z$} ;

        \draw[thick, red, name path=Wl](-3, -2.65) -- (-2, 0) coordinate(BASE) -- (-3, 2.65) coordinate(TOP) ;
        \draw[thick, red, name path=Wr] (3, -2.65) -- (2, 0) -- (3, 2.65) ;
        
        \draw[dashed] (0, 0) -- (-3, 2.65) ;
        \draw[dashed] (0, 0) -- (-3, -2.65) ;
        \draw[dashed] (0, 0) -- (3, 2.65) ;
        \draw[dashed] (0, 0) -- (3, -2.65) ;
        
        \draw[red, name path=T1] (-3, 2.65) .. controls (-1, 2.2) and (1, 2.2) .. (3, 2.65) ;
        \draw[red, dashed, name path=T2] (-3, 2.65) ..controls(-1, 3) and (1, 3) .. (3, 2.65) ;
        
        \draw[red, name path=B1] (-3, -2.65) .. controls (-1, -3.2) and (1, -3.2) .. (3, -2.65) ;
        \draw[red, dashed, name path=B2] (-3, -2.65) ..controls (-1, -2.3) and (1, -2.3) .. (3, -2.65) ;
        
        \pic [draw, <-, "$\alpha_{l}$", angle eccentricity=1.5] {angle=TOP--BASE--D_out} ;
        \pic [draw, ->, "$\theta_{m}$", angle eccentricity=1.5] {angle=Z--BH--TOP} ;

        \tikzfillbetween[of=T1 and T2] {red, opacity=0.1} ;
        \tikzfillbetween[of=T1 and B1] {red, opacity=0.2} ;
        
        \draw[color=white, name path=lf, opacity=0] (-3,-2.65) -- (-3, 2.65) ;
        \draw[name path=cl, opacity=0] (-3, 2.65) arc [start angle=138.5, delta angle=80, radius=4] ;
        \tikzfillbetween[of=lf and cl] {red, opacity=0.2} ;
        
        \draw[name path=rf, opacity=0] (3, -2.65) -- (3, 2.65) ;
        \draw[color=blue, name path=cr, opacity=0] (3, -2.65) arc [start angle=-41.5, delta angle=80, radius=4] ;
        \tikzfillbetween[of=rf and cr]{red, opacity=0.2} ;
        
        \draw[black] (-2, -0.5) -- (-2, 0.1) ;
        \node (rl) at (-1.85, -0.65) {$r_{l}$} ;
        
        \draw[dash dot, black] (-3, 0) -- (-3, 2.65) ;
        \node (r) at (-3, -0.2) {$r_{\mathrm{max}}$} ;
        \node (h) at (-3.2, 2.9) {$h_{\mathrm{max}}$} ;
        
    \end{tikzpicture}
    \caption{A sketch of the geometry used for the bi-conical outflow model. The outflows are shown as the solid red lines, travelling from the disc (green) to the outer edge of the sphere. The transparent red region shows the area of the sky this outflow subtends, as seen by the BH. $\alpha_{l}$ is the wind launch angle, $r_{l}$ is the wind launch radius, while $r_{\mathrm{max}}$ and $h_{\mathrm{max}}$ are the radius and height of the top of the wind, while $\theta_{m}$ is the corresponding polar angle. The covering fraction given as a model parameter is then simply the solid angle of the red band divided by $4\pi$.}
    \label{fig:wind_solid}
\end{figure}

Firstly, we need to determine the limits of our wind surface. Since we assume it takes the shape of a bi-cone, it will be launched from all azimuths. Hence $\phi$ ranges from $0$ to $2\pi$. As the wind is launched from the disc in the x-y plane $\cos(\theta)$ will range from $0$ to $\cos(\theta_{m})$, where $\theta_{m}$ is the polar angle for the top of the wind (i.e the maximal extent of the outflow). The wind will region in the sky (as seen from the BH) illustrated by the red band in Fig. \ref{fig:wind_solid}. Hence we can relate $\theta_{m}$ to $f_{\mathrm{cov}}$ through the solid angle of this red band, since $f_{\mathrm{cov}} = \Omega_{\mathrm{band}}/4\pi$. From the definition of the solid angle we have $d\Omega = \sin(\theta) d\theta d\phi$. Since $\theta$ is measured from the z-axis and down it is simpler to calculate the solid angle of the top conical section, $\Omega_{\mathrm{con}}$ (non-shaded regions in Fig. \ref{fig:wind_solid}), and relate it to that of the red band through $\Omega_{\mathrm{band}} = 4\pi - 2\Omega_{\mathrm{con}}$. This gives us the following expression for the band solid angle

\begin{equation}
    \frac{\Omega_{\text{band}}}{4\pi} = \cos(\theta_{m}) = f_{\mathrm{cov}}
\end{equation}

We now divide our wind area into a polar grid in $\cos(\theta)$ and $\phi$, where $\phi$ is linearly spaced between $0$ and $2\pi$ with spacing $d\phi$, and $\cos(\theta)$ is linearly spaced between $0$ and $\cos(\theta_{\text{m}})$ with spacing $d\cos(\theta)$.

Since our wind is now fully described by the polar coordinates $\theta$ and $\phi$, the solid angle subtended by each grid-point is $d\Omega_{\text{grid}} = d\cos(\theta) d\phi$. The X-ray luminosity seen at each grid-point is then $L_{x} (d\Omega_{\text{grid}}/4\pi)$, giving the re-processed luminosity  at each grid point as

\begin{equation}
    dL_{\text{rep}} = (1-A_{w})L_{x} \frac{d\cos(\theta) d\phi}{4\pi}
\end{equation}

where $A_{w}$ is the wind albedo.

\begin{figure}
    \centering
    \begin{tikzpicture}
        
        \filldraw[black] (6, 0) circle (4pt) ;
        \filldraw[blue] (6, 1.5) circle (2pt) ;
        \draw[green, very thick] (5.5, 0) -- (0, 0) coordinate(Outer) ;
        \draw[red, very thick] (1, 0) coordinate(Base) -- (0, 3) coordinate(Top) ;
        
        \pic [draw, <-, "$\alpha_{l}$", angle eccentricity=1.5] {angle=Top--Base--Outer} ;
        
        \draw[->] (6, 0) -- (6, 1.43);
        \draw[->] (6, 0) -- (0.4, 2) ;
        \draw[->, dotted] (0.4, 2) -- (7, 4) ;
        \draw[->] (6, 1.5) -- (0.4, 2) ;
        \draw[->] (6, 1.5) -- (7, 1.8) ;
        \draw[->] (0.4, 2) -- (5.4, 3.5) ;
        
        \draw[dashed, ->] (6, 1.42) -- (5.4, 3.5) ;
        
        \node (rvec) at (3, 0.8) {$\mathbf{\underline{r}}$} ;
        \node (lvec) at (3.8, 1.45) {$\mathbf{\underline{l}}$} ;
        \node (ldvec) at (3.5, 3.15) {$\mathbf{\underline{l}'}$} ;
        \node (svec) at (5.9, 2.5) {$\mathbf{\underline{s}}$} ;
        \node (ivec) at (7.4, 1.85) {$\mathbf{\underline{\hat{i}}_{\text{obs}}}$} ;
        \node (hx) at (6.25, 0.7) {$h_{x}$} ;
        
    \end{tikzpicture}
    \caption{A sketch of the geometry used to determine the time delay for each point on the wind surface. The blue circle indicates the position of the X-ray source in the lamppost approximation, and the vector $\mathbf{\underline{\hat{i}}_{\text{obs}}}$ is in the direction of the observer.}
    \label{fig:wind_delay}
\end{figure}
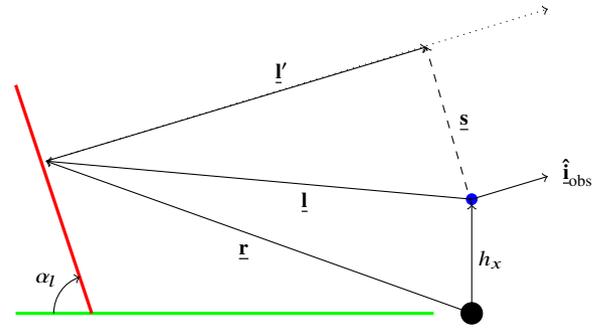

However, to understand the response we still need to know the time-delay at each grid-point. Again, we follow the method of \citet{Welsh91}, but for the geometry sketched in Fig. \ref{fig:wind_delay}. The time delay is then simply the light-travel time over the path difference between the direct emission, and the emission passing via the wind: $\tau = (R_{G}/c) (|\mathbf{\underline{l}} + \mathbf{\underline{l}'}|)$. We can relate $\mathbf{\underline{l}}$ to the positional coordinates $(r, \phi, h)$ through.

\begin{equation}
    \mathbf{\underline{l}} =
    \begin{pmatrix}
        r \cos(\phi) \\
        r \sin(\phi) \\
        h - h_{x} 
    \end{pmatrix}
    \label{eqn:lvec}
\end{equation}

where $h_{x}$ is the height of the lamppost corona, and $\phi$ is measured from the x-axis within the x-y plane. For $\mathbf{\underline{l}'}$ we note that the path difference only extends to the point where $\mathbf{\underline{l}'}$ is tangential to the vector from the X-ray source to its tip, $\mathbf{\underline{s}}$. Since $\mathbf{\underline{l}'}$ must necessarily travel in the direction of the observer, we can also write $\mathbf{\underline{l}'} = d' \mathbf{\underline{\hat{i}}_{\text{obs}}}$, where

\begin{equation}
    \mathbf{\underline{\hat{i}}_{\text{obs}}} =
    \begin{pmatrix}
         \sin(i) \\
         0 \\
         \cos(i) 
    \end{pmatrix}
\end{equation}

is the inclination vector. Writing $\mathbf{\underline{s}} = \mathbf{\underline{l}} + \mathbf{\underline{l}'}$, setting $\mathbf{\underline{l}'} \cdot \mathbf{\underline{s}} = 0$, and solving for $d'$, we have

\begin{equation}
    d' = (h_{x} - h) \cos(i) - r \sin(i) \cos(\phi)    
\end{equation}

which gives us a time-delay

\begin{equation}
    \tau(r, \phi) = \frac{R_{G}}{c} = \bigg\{ \sqrt{r^{2} + (h-h_{x})^{2}} + (h_{x}-h) \cos(i) - r\sin(i) \cos(\phi) \bigg\}
\end{equation}

However, as we have defined our grid in terms of $\cos(\theta)$ and $\phi$, we need to transform our radial and vertical coordinates such that

\begin{equation}
    r(\theta) = \frac{r_{l} \tan(\alpha_{l})}{\tan(\alpha_{l}) - \tan(\frac{\pi}{2} - \theta)}
\end{equation}

\begin{equation}
    h(\theta) = r(\theta) \tan \left(\frac{\pi}{2} - \theta \right)
\end{equation}

Finally, putting this all together, and making the assumption that $h_{x}$ is sufficiently small that the solid angle of a grid point seen from the BH is the same as that seen from the X-ray source, we have that the luminosity of each point must vary as

\begin{equation}
    dL_{\text{rep}}(t, \theta, \phi) = (1-A_{w}) L_{x}\big(t-\tau(r(\theta), \phi) \big) \frac{d\cos(\theta) d\phi}{4\pi}
\end{equation}

and so the total outflow luminosity varies as

\begin{equation}
    L_{\text{rep}}(t) = \sum_{\theta=\frac{\pi}{2}}^{\theta_{\text{m}}} \sum_{\phi=0}^{2\pi} dL_{\text{rep}}(t, \theta, \phi)
\end{equation}

\section{Varying the Outflow Parameters}

\begin{figure}
    \centering
    \includegraphics[width=\columnwidth]{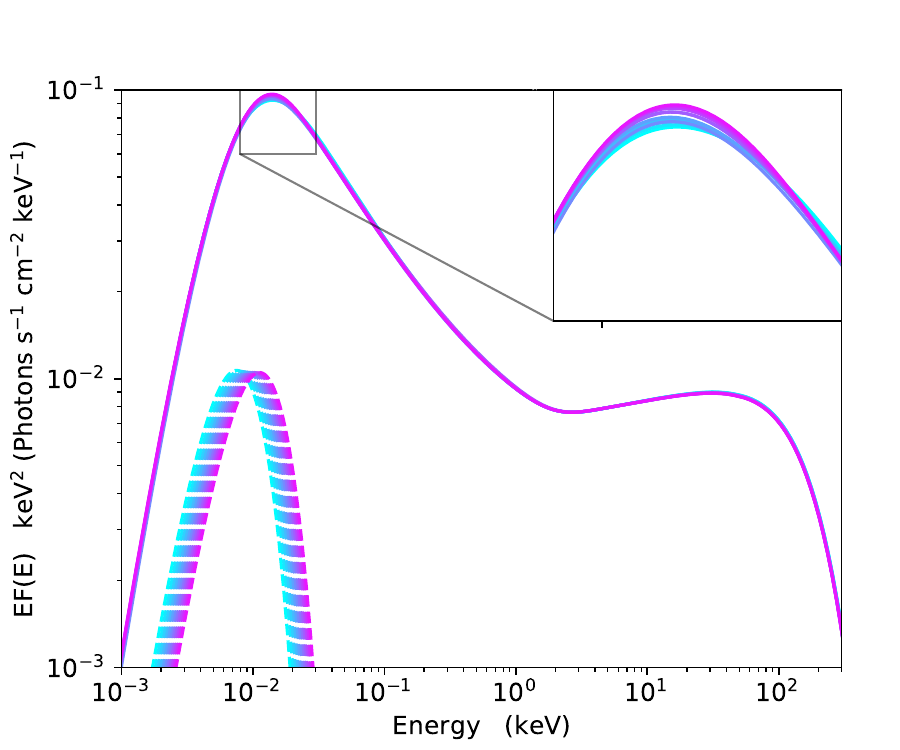}
    \caption{The set of SEDs used for the parameter scan in the outflow model. These have been calculated by fitting {\sc agnref} to the spectral data over the range of $kT_{\text{out}}$. The colour-scheme is ascending from blue throught to purple as we increase $kT_{\text{out}}$. The solid lines show the total SEDs, while the dashed lines show the thermal, diffuse, components. Note that we have removed the reflected component for clarity, as we are mostly interested in the intrinsic SED and thermal component when considering the UV variability. The inset shows a zoomed version of the SED peaks, as this is where we see the greatest effect of changing $kT_{\text{out}}$.}
    \label{fig:BiCon_SEDscan}
\end{figure}

\begin{figure*}
    \centering
    \includegraphics[width=\textwidth]{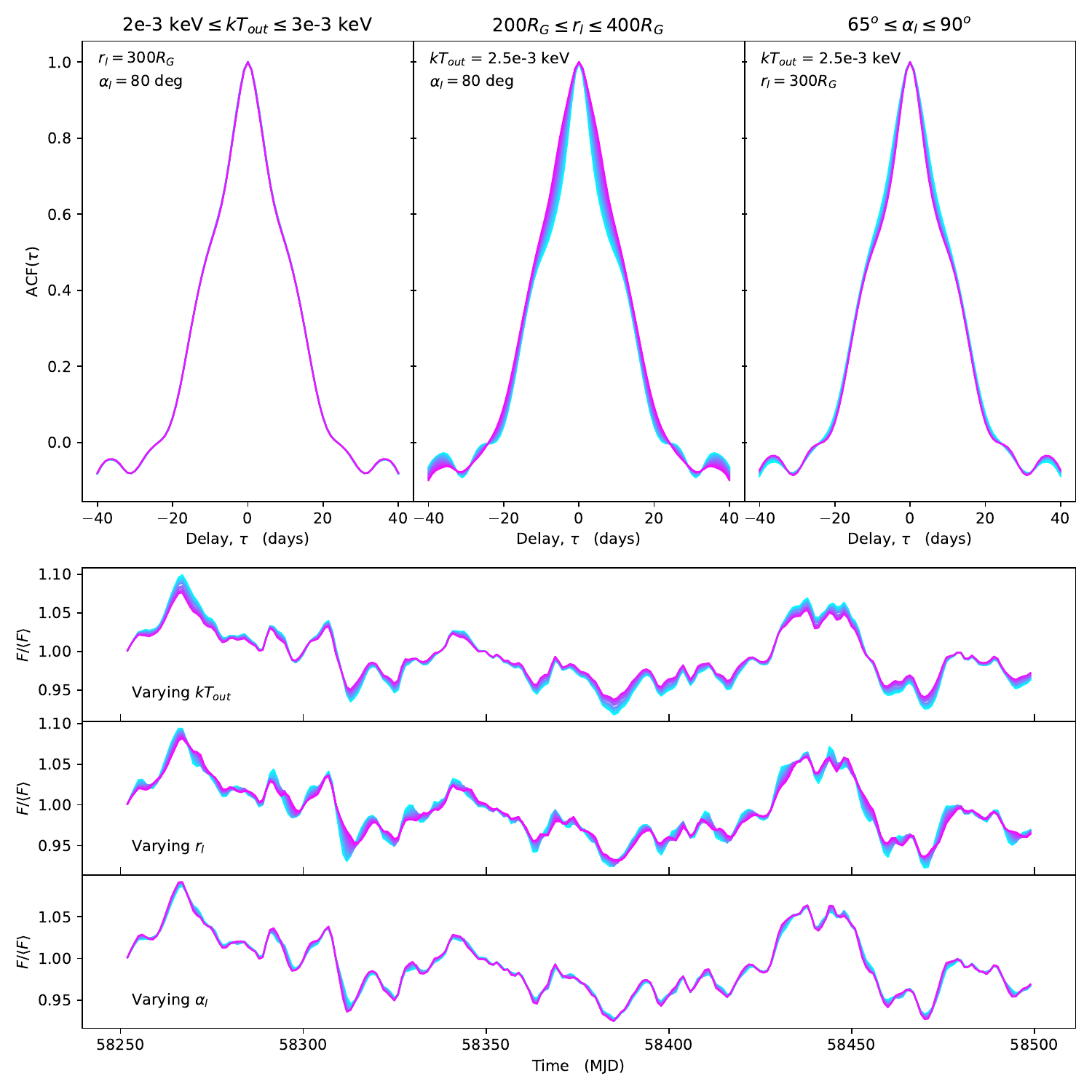}
    \caption{Model light-curves from the {\sc agnref} model. The top row show the ACFs, while varying $kT_{\text{out}}$ (left), $r_{l}$ (middle), and $\alpha_{l}$ (right). The bottom three rows show the model light-curves, again while varying $kT_{\text{out}}$ (top), $r_{l}$ (middle), and $\alpha_{l}$ (bottom). The colour-scheme is set such that it ascends from blue to purple as the relevant parameter increases in value (i.e the lowest parameter value gives a light blue line, while the highest gives a deep purple line). While not being varied, each parameter is kept constant at $kT_{\text{out}} = 2.5\times10^{-3}$\,keV, $r_{l}=300\,R_{G}$, and $\alpha_{l}=80$\,deg.}
    \label{fig:BiCon_paramScan}
\end{figure*}

Here we asses the impact of the outflow black-body temperature, $kT_{\text{out}}$, launch radius, $r_{l}$, and launch angle, $\alpha_{l}$ (measured from the disc), on the output model-light-curves. We vary $kT_{\text{out}}$ from $2\times 10^{-3}$\,keV to $3\times10^{-3}$\,keV, $r_l$ from $200\,R_{G}$ to $400\,R_{G}$, and $\alpha_{l}$ from 65\,deg to 90\,deg. For each temperature $kT_{\text{out}}$ we construct an SED using our {\sc agnref} model, in order to constrain the energetics and outflow covering fraction, before we calculate the model light-curves. Hence, for each value of $kT_{\text{out}}$ we use a slightly different SED to generate the light-curves. These are shown in Fig. \ref{fig:BiCon_SEDscan}. Additionally, we only consider launch angles $\alpha_{l} \geq 65$\,deg, since for certain combinations of $r_{l}$ and $f_{\text{cov}}$ (predicted from the SED) it is not possible to reach the correct solid angle for $\alpha_{l} \lesssim 65$\,deg. These results are shown in Fig. \ref{fig:BiCon_paramScan}.

It is clear from Fig. \ref{fig:BiCon_SEDscan} that varying $kT_{\text{out}}$ has only a marginal effect on the total SED. It does, however, have a notable effect on the response within the light-curve, where we can see that reducing $kT_{\text{out}}$ within the limits we have set gives an increase in the variability amplitude. This is simply explained that as we decrease the temperature of the thermal component, it becomes more dominant in the SED at energies associated with the UVW2 emission. Additionally, changing the outflow temperature makes no difference in how much X-ray flux the outflow sees, and hence no difference in the total variability of the thermal component. Hence, when we shift the thermal emission to a temperature where it makes a greater contribution to the band we are observing in we will see a higher variability amplitude; even though the overall variability of the component itself is not changing. This also explains why the ACF does not change when we vary $kT_{\text{out}}$, since the time-scales the variations occur on are also not changing.

Moving onto the launch radius we see that changing $r_{l}$ has an effect on both the light-curves and the ACF. This is important, since $r_{l}$ cannot be constrained by the SED, hence we can only use temporal information (i.e the light-curves, ACF, or CCF) to estimate this parameter. Firstly, we note that increasing $r_{l}$ increases the smoothing effect in the light-curve, as highlighted by the widening of the narrow component in the ACF. This is entirely expected, as increasing $r_{l}$ will increase the light-travel time to the outflow, which in turn increases the time-scale over which smoothing occurs. We also see slight reductions in amplitude around some of the sharper peaks in the light-curve, as $r_{l}$ increases. This is an effect of smoothing, not geometry; since the covering fraction is kept constant for each value of $r_{l}$, and so the outflow sees the same fraction of X-ray luminosity no matter the launch radius. Smoothing causes this apparent reduction in amplitude because an increase in smoothing leads to a marginalisation over a greater range of X-ray luminosities within the input light-curve for each time-step in the model light-curve. This explains why we only see this reduction around the sharp peaks, as these will be most strongly affected by the increased range in X-ray luminosities. In other words, the smoothing works exactly as one would expect it to. 

Finally, we examine the effect of varying the launch angle $\alpha_{l}$. Firstly, we note that there appears to be a narrower range in output light-curves and ACFs when varying $\alpha_{l}$ compared to $r_{l}$ and $kT_{\text{out}}$. This is most likely due to the limit range in $\alpha_{l}$ that we are exploring. We can see, however, that decreasing the launch angle does have an effect on the smoothing of the light-curve, as we can see a slight widening of the narrow core in the ACF when $\alpha_{l}$ is reduced. This is an effect arising from the way our model hard-wires the outflow solid angle. $f_{\text{cov}}$ is set by the SED, and so remains constant under variations in $\alpha_{l}$. This leads to the radial extent of the outflow increasing as $\alpha_{l}$ decreases, since the outflow needs to extend to sufficient radii such that it reaches a height large enough to satisfy the solid angle set by $f_{\text{cov}}$. Clearily increasing the radial extent of the outflow will increase the light-travel time, and so increase the range of time-delays across the outflow grid; finally leading to an increased smoothing effect.

\newpage

\section{The response from a Comptonised Disc}

\begin{figure*}
    \centering
    \includegraphics[width=\textwidth]{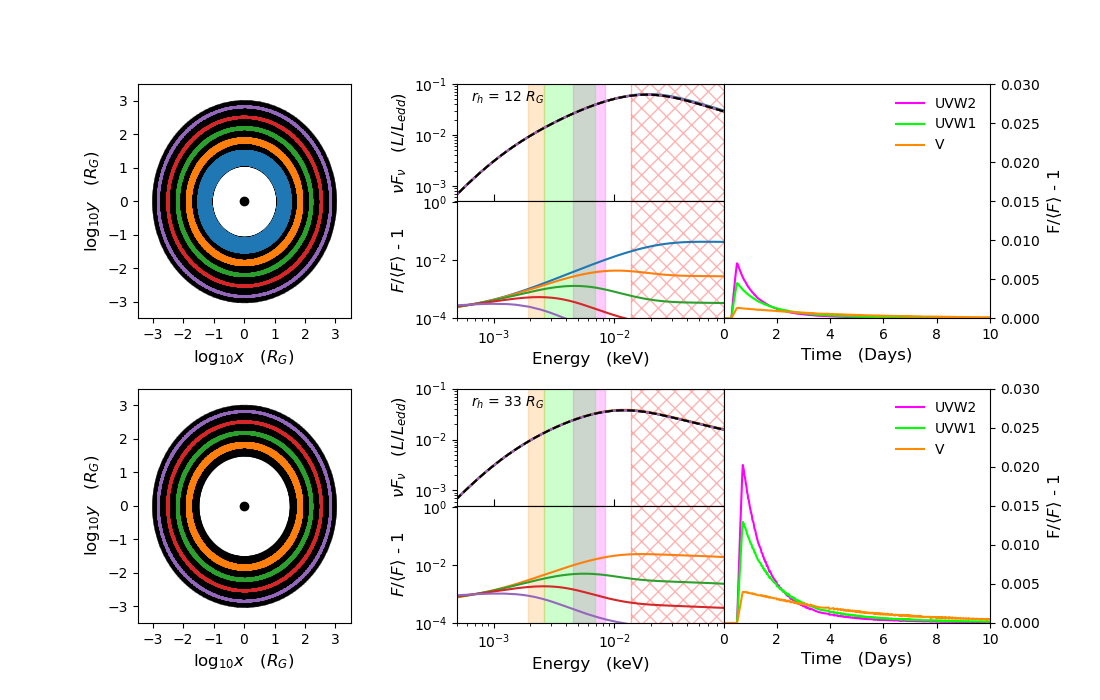}
    \caption{Same as Fig. \ref{fig:sedsnaps_disc}, but for a Comptonised disc}
    \label{fig:SEDsnaps_warm}
\end{figure*}

Repeating the experiment from section 2.4, but for a Comptonised disc, we find that although there is a change in spectral shape, the response functions remain almost identical. This is because the spectrum still strongly peaked at an energy linearly related to the seed photon temperature from the disc as the Comptonisation is steep. 
The results are shown in Fig. \ref{fig:SEDsnaps_warm}. Clearly, like the case for a standard disc, the response at all energies is dominated by the inner disc response. 

\section{Swift, NICER, or XMM}

\begin{figure}
    \centering
    \includegraphics[width=\columnwidth]{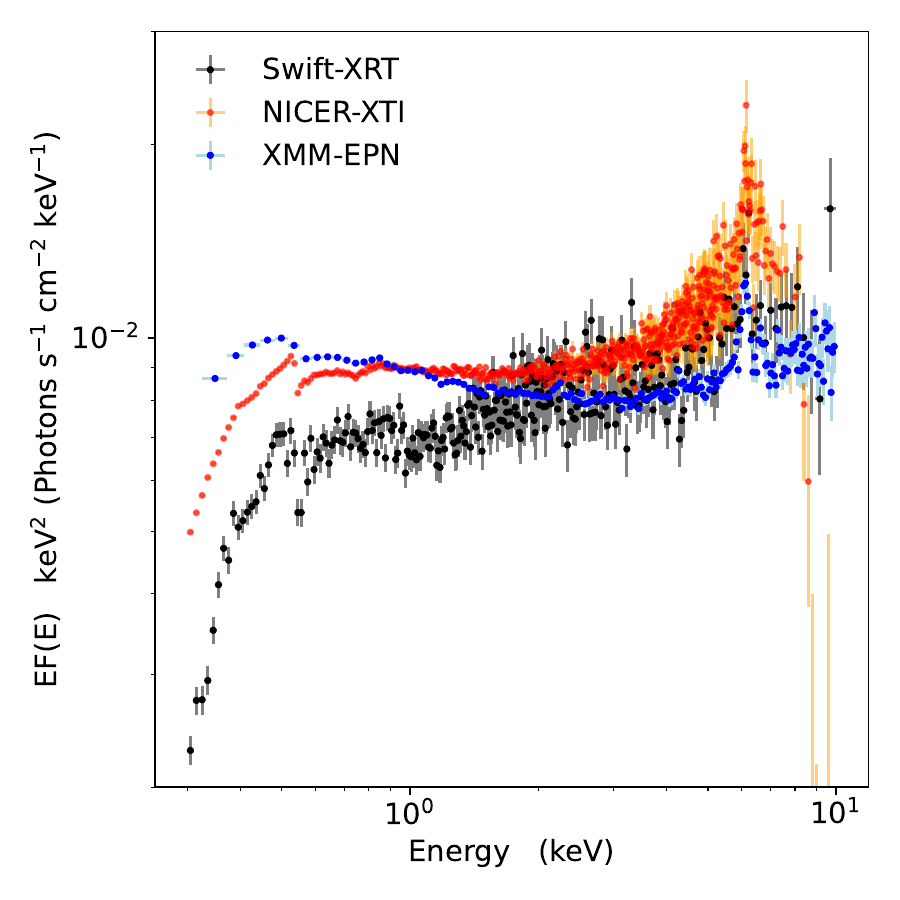}
    \caption{A comparison of Swift-XRT data (black), NICER-XTI data (red) taken during the observation campaign, and archival XMM-EPN data (blue).}
    \label{fig:sw_v_ni_v_xm}
\end{figure}

In section 3.2 we made the choice of using the archival 
XMM-EPN data instead of the campaign data from NICER-XTI or
Swift-XRT. Fig. \ref{fig:sw_v_ni_v_xm} shows a comparison of
these spectral data. The Swift (black) and NICER (red) data
are surprisingly quite different below 2~keV, with Swift being 
$\sim 30$\,\% dimmer at the lowest energies. We checked that this was not due to the different time sampling. The mean 
count rate in the Swift-XRT light-curve across the entire monitoring period is $\sim 0.86$\,counts\,s$^{-1}$ which is actually slightly higher than the count rate in Swift-XRT for periods corresponding to just the NICER observation periods ($\sim 0.83$\,counts\,s$^{-1}$). Thus the difference must be due to cross-calibration uncertainties rather than intrinsic to F9. It is not clear which of NICER or Swift is closer to the 'real' spectrum, though with Swift being the older (and less sensitive) instrument then it is perhaps more likely that this has been subject to degradation/contamination at low energies. Using NICER would then then be the obvious choice, except that this instrument has systematic uncertainties at higher energies due to complexity of background subtraction. 
The derived spectrum is very dependent on the background subtraction for $E\gtrsim 5$\,keV, yet even our 
best attempt, using the newly released {\sc scorpeon} background model (released with v. 6.31 of {\sc heasoft}), over-subtracts background at the highest energies, as the counts go negative for $E \gtrsim 8$\,keV. 

In essence, we do not trust NICER at high energies and we do not trust SWIFT at low energies. The archival XMM-Newton data (blue) give a better match to the NICER data than Swift at low energies, and a better match to Swift than NICER at high energies, so we choose this to set the time averaged spectral shape. 

\bsp	
\label{lastpage}
\end{document}